\documentclass[journal,comsoc]{IEEEtran}
\usepackage[T1]{fontenc}

\usepackage{cite}

\usepackage{amsmath}
\usepackage[cmintegrals]{newtxmath}

\usepackage{algorithm}
\usepackage{algpseudocode}
\usepackage{amsmath}
\usepackage{amssymb}
\usepackage{mathtools}
\usepackage{bm}
\usepackage{tikz}
\usetikzlibrary{shapes,arrows,decorations.markings,
	calc,decorations.pathreplacing,intersections,chains,dsp,
	positioning,arrows,fit,external,arrows.meta}
\usepackage{pgfplots}
\pgfplotsset{compat=1.8}
\usepackage{standalone}
\usepackage{hyperref}
\hypersetup{draft, debug}
\usepackage{pgfplots}
\usepackage{tabulary}

\newtheorem{definition}{Definition}

\usepackage{caption}
\usepackage{subcaption}

\usepackage[shortcuts, acronym, toc,nonumberlist=false]{glossaries}
\usepackage{glossary-mcols}
\usepackage{booktabs}
\makenoidxglossaries
\newacronym{MIMO}{MIMO}{Multiple Input Multiple Output}
\newacronym{SISO}{SISO}{Single Input Single Output}
\newacronym{SIMO}{SIMO}{Single Input Multiple Output}
\newacronym{MISO}{MISO}{Multiple Input Single Output}
\newacronym{SU}{SU}{Single-User}
\newacronym{MU}{MU}{Multi-User}
\newacronym{ADC}{ADC}{Analog-to-Digital Converter}
\newacronym{DAC}{DAC}{Digital-to-Analog Converter}
\newacronym{BS}{BS}{Base Station}
\newacronym{SNR}{SNR}{Signal-to-Noise Ratio}
\newacronym{wrt}{w.r.t.}{with respect to}
\newacronym{iff}{iff}{if and only if}
\newacronym{flops}{FLOPs}{floating-point operations}
\newacronym{AWGN}{AWGN}{Additive White Gaussian Noise}
\newacronym{iid}{i.i.d.}{independent and identically distributed}
\newacronym{MUI}{MUI}{Multi-User Interference}
\newacronym{QE}{QE}{Quantization Error}
\newacronym{SQINR}{SQINR}{Signal-to-Quantization-plus-Interference-plus-Noise Ratio}
\newacronym{WL}{WL}{Widely-Linear}
\newacronym{SL}{SL}{Strictly-Linear}
\newacronym{OFDM}{OFDM}{Orthogonal Frequency-Division Multiplexing}
\newacronym{5G}{5G}{5$ ^{\text{th}} $ Generation Wireless Systems}
\newacronym{RF}{RF}{Radio Frequency}
\newacronym{PA}{PA}{Power Amplifier}
\newacronym{CSI}{CSI}{Channel State Information}
\newacronym{MF}{MF}{Matched Filter}
\newacronym{MMSE}{MMSE}{Minimum Mean Squared Error}
\newacronym{MSE}{MSE}{Mean Squared Error}
\newacronym{BER}{BER}{Bit Error Rate}
\newacronym{TxWF}{TxWF}{Transmit Wiener Filter}
\newacronym{bpcu}{bpcu}{bit(s) per channel use}
\newacronym{ZF}{ZF}{Zero Forcing}
\newacronym{TxWFQ}{TxWFQ}{Quantized Transmit Wiener Filter}
\newacronym{QAM}{QAM}{Quadrature Amplitude Modulation}
\newacronym{QPSK}{QPSK}{Quadrature Phase Shift Keying}
\newacronym{PSK}{PSK}{Phase Shift Keying}

\newcommand{\matt}[1]{\bm{#1}}
\newcommand{\vect}[1]{\bm{#1}}
\newcommand{\abs}[1]{\left \lvert{#1} \right \rvert}

\newcommand{\T}{\text{T}}

\DeclareMathOperator*{\argmin}{arg\,min}
\DeclareMathOperator*{\argmax}{arg\,max}
\DeclareMathOperator*{\E}{\mathrm{E}}
\newcommand{\rank}[1]{\mathrm{rank}\left ({#1}\right)}
\newcommand{\diag}[1]{\mathrm{diag}\left ({#1}\right)}
\newcommand{\nondiag}[1]{\mathrm{nondiag}\left ({#1}\right)}
\newcommand{\tr}[1]{\mathrm{tr}\left ({#1}\right)}

\newcommand{\Nt}{{N}_{\mathrm{t}}}

\newcommand{\jim}{\mathrm{j}\,}
\DeclareMathOperator*{\ccal}{\mathcal{C}}
\DeclareMathOperator*{\ncal}{\mathcal{N}}
\DeclareMathOperator*{\qcal}{\mathcal{Q}}
\DeclareMathOperator*{\pcal}{\mathcal{P}}
\newcommand*{\st}{\mathrm{\quad s.t.\quad}}
\newcommand{\figref}[1]{\text{Fig.}~\ref{#1}}

\begin{document}
%
\title{Reconsidering Linear Transmit Signal Processing in 1-Bit Quantized Multi-User MISO Systems}
%
\author{Oliver~De~Candido,~\IEEEmembership{Student~Member,~IEEE,}
        Hela~Jedda,~\IEEEmembership{Student~Member,~IEEE,}
        Amine~Mezghani,~\IEEEmembership{Member,~IEEE,}
        A.~Lee~Swindlehurst,~\IEEEmembership{Fellow,~IEEE,}
        and~Josef~A.~Nossek,~\IEEEmembership{Life~Fellow,~IEEE}
\thanks{O. De Candido, H. Jedda and J. A. Nossek are with the Associate Professorship of Signal Processing, Department of Electrical and Computer Engineering, Technische Universit\"at M\"unchen, 80290 Munich, Germany (e-mail: \{oliver.de-candido, hela.jedda, josef.a.nossek\}@tum.de).}
\thanks{A. Mezghani is with the Wireless Networking and Communications Group, University of Texas at Austin, Austin, TX 78712, USA (e-mail: amine.mezghani@utexas.edu).}
\thanks{A. L. Swindlehurst is, and A. Mezghani was, with the Center for Pervasive Communications and Computing, University of California Irvine, Irvine, CA 92697, USA (e-mail: swindle@uci.edu).}
\thanks{J. A. Nossek is also with the Department of Teleinformatics Engineering, Universidade Federal do Cear\'a, Fortaleza 60020-181, Brasil.}
\thanks{Manuscript submitted to IEEE Transactions on Wireless Communications on February 27, 2018}} %

\maketitle

\begin{abstract}
	In this contribution, we investigate a coarsely quantized \gls{MU}-\gls{MISO} downlink communication system, where we assume 1-Bit \glspl{DAC} at the \gls{BS} antennas.  First, we analyze the achievable sum rate lower-bound using the Bussgang decomposition.  In the presence of the non-linear quantization, our analysis indicates the potential merit of reconsidering traditional signal processing techniques in coarsely quantized systems, i.e., reconsidering transmit covariance matrices whose rank is equal to the rank of the channel.  Furthermore, in the second part of this paper, we propose a linear precoder design which achieves the predicted increase in performance compared with a state of the art linear precoder design.  Moreover, our linear signal processing algorithm allows for higher-order modulation schemes to be employed.
\end{abstract}

\glsresetall
\IEEEpeerreviewmaketitle


\section{Introduction and Motivation} \label{sec:Intro_Motivation}
\IEEEPARstart{I}{n} recent years, the demand for higher data rates has drastically increased as the number of personal and inter-connected devices continuously increases (e.g., Internet of Things).  These demands should be fulfilled by \gls{5G}  under similar cost and energy constraints as current wireless communication systems.  To this end, two complimentary technologies have been introduced at the forefront of research to provide the required data rates for \gls{5G}.  First, the use of a large number of antennas at the \gls{BS}, referred to as massive \gls{MIMO}, has been investigated.  Due to the inherent high antenna diversity and array gain, these systems have shown improvements in data throughput, spectral and radiated energy efficiency, using relatively simple processing, see e.g., \cite{larsson2014massive, lu2014overview, hoydis2013massive}.  Second is the use of millimeter wave (mmWave) carrier frequencies, where the large amount of available bandwidth will allow for higher data rates, e.g., \cite{hong2014study, rappaport2013millimeter}.

If future communications systems are equipped with many \gls{BS} antennas (hundreds or even thousands), and/or are working at higher sampling rates, the requirement for power and cost efficient components in the \gls{RF} chain at each antenna is evident.  Currently, the most power hungry component in the \gls{RF} chains are the \glspl{PA}, \cite{blume2010approaches,chen2010energy}.  \glspl{PA} are most energy efficient when operated in their saturation region; however, in this region, they introduce non-linear distortions to the transmit signal. These distortions can be avoided when constant envelope input signals are employed, i.e., signals with a constant magnitude, and thus such amplitude distortions can be ignored.  Furthermore, the power consumption of the \glspl{DAC}/\glspl{ADC} in the \gls{RF} chains increases exponentially with their resolution (in bits) and linearly with the sampling frequency, i.e., $ {P_{\mathrm{diss}} \propto 2^{b}\cdot f_{s}} $, \cite{murmann2015adc,svensson2006power,walden1999analog}.  Thus, a simple solution to reduce power consumption and chip area, whilst simultaneously employing constant envelope modulation, is to use low-resolution (or coarsely-quantized) \glspl{DAC}/\glspl{ADC}.  In this paper, we focus on the downlink scenario with the coarsest form of quantization, i.e., systems where the \gls{BS} is equipped with 1-bit \glspl{DAC}.

It has been shown, see e.g., \cite{krone2012capacity,koch2010icreased,jacobsson2017massive}, that systems employing oversampling at the transmitters/receivers can improve the performance limitations introduced by the 1-bit \glspl{DAC}/\glspl{ADC}.  Furthermore, the issue of spectral shaping with 1-bit \glspl{DAC} and oversampling was investigated in \cite{jedda2015spectral}, where it was shown that despite the low-resolution quantization, sufficient spectral confinement can be achieved.  As we only consider spatial filtering and discrete-time processing, we focus on symbol-sampled models.


\subsection{Existing Work} \label{sec:Intro_Sub_Existing_Work}
Recent research into the topic of coarsely-quantized \gls{MIMO} systems can be categorized as focusing on either the uplink or the downlink scenario, where the \gls{BS} is assumed to have low-resolution \glspl{ADC} or \glspl{DAC}, respectively.


\subsubsection{Uplink}\label{sec:Intro_Sub_Existing_Work_UL}
The capacity of coarsely-quantized \gls{MIMO} systems was originally investigated in \cite{nossek2006capacity}, which showed only a small loss in capacity comparing quantized and unquantized \gls{MIMO} systems. However, \cite{nossek2006capacity} and \cite{ivrlac2006challenges} show that coding becomes an issue, since traditional channel coding methods are unsuitable for quantized \gls{MIMO} systems.  In \cite{mezghani2007ultra}, the Taylor expansion of the mutual information up to the second-order is derived, which shows a $ 2/\pi $ loss in achievable rate at low-\gls{SNR}.  This loss, due to the use of symmetric threshold quantizers, was also reported in \cite{koch2013at}.  Moreover, a mutual information lower-bound was derived in \cite{Mezghani2012capacity}, based on the Bussgang theorem \cite{Bussgang1952}, which confirms the $ 2/\pi $ loss at low-\gls{SNR}. 

In \cite{mo2014high} and \cite{mo2015capacity}, a closed-form expression for the capacity of the \gls{SISO} and \gls{MISO} uplink scenarios is derived, assuming perfect \gls{CSI}.  Moreover, capacity bounds are found for the general \gls{MIMO} scenario, and the mutual information lower-bound from \cite{Mezghani2012capacity} was shown to be tight at low-\gls{SNR} but loose at high-\gls{SNR}. Furthermore, \cite{jacobsson2015one} shows that higher-order modulation is possible with 1-bit quantized \glspl{ADC}.


\subsubsection{Downlink}\label{sec:Intro_Sub_Existing_Work_DL}
A lower-bound for the achievable rate in quantized \gls{MIMO} systems was derived in \cite{li2017downlink}, assuming matched filter precoding and estimated \gls{CSI}.  Moreover, in \cite{li2017downlink}, for single-antenna users, it was shown that roughly 2.5 times more \gls{BS} antennas are required to achieve the same rates as in unquantized systems for maximum ratio precoding.  In \cite{decandido2017are}, the validity of traditional signal processing techniques was questioned for quantized single-user \gls{MISO} systems, i.e., whether proper signaling and transmit covariance matrices whose rank is equal the rank of the channel matrix (channel rank) are still optimal in the presence of the non-linear quantization.

Recent research has also focused on linear and non-linear transmit signal processing techniques in quantized \gls{MIMO} downlink systems; one of the first linear signal processing designs taking quantization into account was introduced in \cite{Mezghani2009transmit}.  Therein, a \gls{TxWFQ} was designed using the optimal quantization step-size and linearizing the quantization operation.  In \cite{usman2016mmse}, a linear precoder and an analog power allocation matrix were designed to minimize the \gls{MSE} using a gradient projection algorithm.  It should be noted that a precoder design using the optimal quantization step-size (e.g., \cite{Mezghani2009transmit}) is equivalent to using constant step-sizes and introducing an analog real-valued diagonal power allocation matrix (e.g., \cite{usman2016mmse}).  A linear precoder designed to maximize the weighted sum rate in a \gls{MU}-\gls{MISO} system was introduced in \cite{kakkavas2016weighted}, where the weighted sum rate is derived using a lower-bound on the achievable rate similar to \cite{Mezghani2012capacity}.  In  \cite{saxena2016analysis} an asymptotic analysis of \gls{MIMO} scenarios is provided where the number of antennas and users increase to infinity.  Moreover, \cite{saxena2016analysis} employs the \gls{ZF} precoder as a benchmark; an asymptotic achievable rate lower-bound is provided based on the Bussgang decomposition \cite{Bussgang1952}, and the authors show that reasonable performance can be obtained if the ratio of antennas to users is large enough.  Applying simple perturbations to the solutions obtained by standard quantized linear precoders has also been shown to improve system performance in \cite{swindlehurst2017minimum}. 

Non-linear precoders, which map the source symbols to the transmit vector in a general way, outperform linear precoders whose outputs are simply truncated by the one-bit quantization, however this comes at the price of higher computational complexity when designing the precoder.  The first non-linear precoder design for low resolution quantized \gls{MIMO} systems was introduced in \cite{mezghani2008tomlinson}, where the Tomlinson-Harashima Precoding method was extended to take the quantization into account.  A novel, non-linear precoder design which optimizes the transmit signal vector by generating lookup-tables for each channel realization was introduced in \cite{jedda2016minimum}.  Furthermore, in \cite{jedda2017massive}, an optimization to reduce the probability of detection error of \gls{PSK} symbols was introduced.  This optimization is based on linear programming, and significantly reduces complexity compared to the lookup-table optimization in \cite{jedda2016minimum}. In \cite{jacobsson2016quantized}, linear and non-linear precoding methods are investigated; it is shown that linear precoding methods only require 3 or 4 bit \glspl{DAC} to achieve performance similar to unquantized systems.  Furthermore, three different non-linear algorithms which minimize the squared error are introduced, and only show a 3 dB loss compared with unquantized systems. 

In \cite{casteneda20171-bit} two non-linear precoder designs based on a biconvex relaxation of the \gls{MSE} minimization is introduced, whereby the second algorithm is optimized to be scalable and have low-complexity with increasing number of \gls{BS} antennas.  A multi-step non-linear precoder design was introduced in \cite{tirkkonen2017subset} to reduce complexity for \gls{QPSK} input symbols.  First, a quantized linear precoder is applied, a subset of transmit antennas are selected and an exhaustive search over a subset-codebook is performed to optimize a  criterion similar to \cite{jedda2016minimum}.  A branch-and-bound approach to maximize the minimum distance to the decision boundary at the receivers is introduced in \cite{landau2017branch}.

Finally, the non-linear algorithms described in \cite{jedda2016minimum} and \cite{jacobsson2016quantized} were extended to allow higher-order modulation schemes in \cite{amor201716} and \cite{jacobsson2016nonlinear}.  In \cite{amor201716}, two lookup-tables are generated per channel realization to allow for a superpositioning of the transmit symbols.  In \cite{jacobsson2016nonlinear}, linear and non-linear algorithms for higher order modulation schemes are introduced, including two algorithms to estimate the receiver scaling factor.  These results show the potential of using higher-order modulation schemes despite the constraint of low-resolution \glspl{DAC}.

The aforementioned non-linear precoder designs optimize the transmit vector symbol-by-symbol at the sampling frequency, which greatly increases their computational complexity.  Therefore, despite the performance gains of the non-linear methods, we are interested in investigating whether linear precoding methods can be improved.


\subsection{Motivation}\label{sec:Intro_Sub_Motivation}
The motivation behind this work stems from the same question we asked in \cite{decandido2017are}: Are traditional signal processing techniques optimal in coarsely quantized \gls{MU}-\gls{MISO} systems?


\subsubsection{Improper Signaling}
First, traditional signal processing techniques often assume that all signals are circular Gaussian distributed, \cite{Adali2011}, i.e., $ s \sim \mathcal{C}\mathcal{N}(0, \sigma_{s}^{2}) $ with $ \sigma_{\Re\lbrace s\rbrace}^{2} + \sigma_{\Im\lbrace s\rbrace}^{2} =  \sigma_{s}^{2}  $, where $ \sigma_{\Re\lbrace s\rbrace}^{2} = \sigma_{\Im\lbrace s\rbrace}^{2} $ represent the variance of the real and imaginary parts, respectively.  Moreover, the real and imaginary parts of $ s $ are assumed to be uncorrelated.  

To motivate the question of whether circular Gaussian signaling is still optimal, we consider the following symmetrical non-convex optimization problem
\begin{align}\label{eq:Example_Motivation}
\min_{x_{1},x_{2}}\lbrace \abs{x_{1}}+\abs{x_{2}}\rbrace \st x_{1}^{2} + x_{2}^{2} = 1.
\end{align}
Despite the fact that \eqref{eq:Example_Motivation} is symmetric \gls{wrt} the variables $ x_{1} $ and $ x_{2} $, i.e., exchanging the variables does not change the objective function nor the constraint, yet the extreme points, $ x_{1,\mathrm{opt}} = \pm 1$ and $ x_{2,\mathrm{opt}} = 0 $ or $ x_{1,\mathrm{opt}} = 0$ and $ x_{2,\mathrm{opt}} = \pm 1 $  are not equal, i.e., $ x_{1,\mathrm{opt}} \neq x_{2,\mathrm{opt}}$.  We can imagine that the constraint in \eqref{eq:Example_Motivation} is the variance of the real and imaginary part of a complex signal $ s $, i.e., $ x_{1} = \sigma_{\Re\lbrace s\rbrace}  $ and $ x_{2} = \sigma_{\Im\lbrace s\rbrace} $.  This would imply that the extreme points allow for unequal power allocation.

In general, optimization problems in quantized \gls{MIMO} systems are non-convex due to the non-linearities and constraints introduced by the quantization.  This simple example, 
of a symmetrical non-convex optimization problem 
motivated us to question the optimality of circular Gaussian distributed signals and proper signaling in quantized \gls{MIMO} systems.


\subsubsection{Higher-Rank Transmit Covariance Matrix}
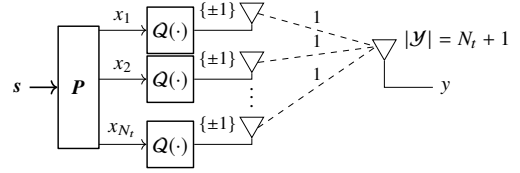
\begin{figure}
	\centering
	\resizebox{0.8\columnwidth}{!}{
		\begin{tikzpicture}
 \tikzstyle{vecArrow} = [thick, decoration={markings,
  	mark=at position 1 with {\arrow[]{angle 90}}},
	double distance=2pt, shorten >= 2.1pt,
	preaction = {decorate},
	postaction = {draw,line width=2pt, white, shorten >= 2.1pt
	}]
	
    \tikzstyle{innerWhite} = [semithick, white,line width=1.4pt, shorten >= 4.5pt]
    
    \tikzstyle{triangle}=[
    isosceles triangle, 
    draw, 
    shape border rotate=-90,
    isosceles triangle apex angle=60,
    inner sep = 0pt,
     minimum height = 1em
    ]
    
    \tikzset{radiation/.style={{decorate,decoration={expanding waves,angle=45,segment length=4pt}}}}

   \tikzset{Bigfilter/.style={shape=rectangle,draw,align=center,text depth=0.3em,text height=1em,inner sep=0pt,
    		line cap=round,line join=round,line width=\dsplinewidth,minimum height=6em,minimum width=2em}}

	\def\stdx{2em}
	\def\stdy{-2em}

	\node[](s) at (0,0) {${\vect{s}} $};
	
	\node[](precoder1a) at ($ (s) + (2*\stdx,-1.4*\stdy) $){};
	\node[](precoder2a) at ($ (s) + (2*\stdx,-0.2*\stdy) $){};
	\node[](precoderNta) at ($ (s) + (2*\stdx,1.4*\stdy) $){};
	
	\node[Bigfilter](P) at ($ (s) +(1.5*\stdx,0)$) {{$ \matt{P} $}};
	
	\node[dspsquare](Q1) at ($ (precoder1a) + (1.75*\stdx,0) $){$ \qcal(\cdot) $};
	\node[dspsquare](Q2) at ($ (precoder2a) + (1.75*\stdx,0) $){$ \qcal(\cdot) $};
	\node[dspsquare](QNt) at ($ (precoderNta) + (1.75*\stdx,0) $){$ \qcal(\cdot) $};
	
	\node[triangle, label=180:{\small$  \lbrace \pm 1\rbrace $}](txAntenna1) at ($ (Q1) + (2*\stdx,-0.5*\stdy) $) []{};
	\node[triangle, label=180:{\small$  \lbrace \pm 1\rbrace $}](txAntennaNt) at ($ (QNt) + (2*\stdx,-0.5*\stdy) $) []{};
	\node[triangle, label=180:{\small$  \lbrace \pm 1\rbrace $}](txAntenna2) at ($ (Q2) + (2*\stdx,-0.5*\stdy) $) []{};
	
	\node[] (tx_dots) at ($ (txAntenna2) + (0,0.8*\stdy) $) {$ \vdots $};
	
	\node[triangle](rxAntenna) at ($ (s) + (9*\stdx,-1*\stdy) $){};
	\node[right] at (rxAntenna.left corner) { {$ \vert \mathcal{Y} \vert  = N_{t} + 1$}};
	
	\node[] (y)	at ($ (s) + (10.5*\stdx,0) $)	{$ {y} $};
	{\draw[->,thick] (s.east) -- (P.west);}
	
	\draw[->] (precoder1a.center) -- (Q1.west)node[above,midway] {${{x}}_{1}$};
	\draw[->] (precoderNta.center) -- (QNt.west)node[above,midway] {${{x}}_{N_{t}}$};
	\draw[->] (precoder2a.center) -- (Q2.west)node[above,midway] {${{x}}_{2}$};

	\draw[] (Q1.east) -| (txAntenna1.apex);
	\draw[] (Q2.east) -| (txAntenna2.apex);
	\draw[] (QNt.east) -| (txAntennaNt.apex);
	
	\draw[dashed] (txAntenna1.left side) -- (rxAntenna.west)node[above,midway] {{\small$1$}};
	\draw[dashed] (txAntenna2.left side) -- (rxAntenna.west)node[above,midway] {{\small$1$}};
	\draw[dashed] (txAntennaNt.left side) -- (rxAntenna.west)node[above,midway] {{\small$1$}};
	
	\draw[] (rxAntenna.apex) |- (y.west); 

\end{tikzpicture}
	}
	\caption{Motivation: Higher-Rank Transmit Covariance Matrix}
	\label{fig:Example_Higher_Rank}
\end{figure}

Second, traditional linear signal processing techniques typically assume that the transmit covariance matrix has the same, or lower, rank than the channel.  As an example, we consider \figref{fig:Example_Higher_Rank}, in which an abstract, real-valued, noiseless quantized single-user \gls{MISO} scenario is depicted, with the rank one channel vector $ \vect{h}^{\T} = [1,\ldots, 1]\in \mathbb{R}^{1\times \Nt} $.  The input signal $ \vect{s} $ is fed into 
the linear precoder matrix $ \matt{P}$, which we assume can be either a vector $ \vect{p} \in \mathbb{R}^{\Nt} $ (a beamforming vector) or a full matrix $ \matt{P} \in \mathbb{R}^{\Nt \times \Nt} $.  Note, we consider linear precoders, the rank of the transmit covariance matrix $ \matt{R}_{\vect{x}} $ is determined by the rank of the precoder matrix.
 The transmit signal is then passed through the 1-bit non-linear quantizers at all transmit antennas; these merely take the sign of the transmit signal, i.e., $ \mathcal{Q}(\vect{x}) : \mathbb{R}^{\Nt} \to \lbrace \pm 1 \rbrace^{\Nt}$. 
 
Next, we consider the signal at the transmit antennas after the 1-bit \glspl{DAC}; here we see that in total we have $ 2^{\Nt} $ distinct transmit signals.  However, if we assume a linear, channel rank precoder vector then the transmit signal is given by: $ \vect{p}\cdot s \in \mathbb{R}^{\Nt} $, where only the sign of $ s $ affects the transmit signal. Thus, we restrict the system to only use 2 of the available $ 2^{\Nt} $ distinct transmit signals.  This implies that the receive constellation yields only two points, $ y \in \mathcal{Y} =  \lbrace \pm \Nt \rbrace $, and the achievable rate is $ I(x;y) \leq 1 $ \gls{bpcu}.
   
If, however, we increase the number of streams available, i.e., $ \vect{s} = [s_{1}, \ldots, s_{\mathcal{R}}] $ with independent symbols $ s_{i} $, and combine them with an augmented precoder matrix $ \matt{P} \in \mathbb{R}^{\Nt \times \mathcal{R}} $ where the columns of $ \matt{P} $ are linearly independent, then we can obtain more distinct transmit signals.  In other words, by increasing the rank of the precoder matrix to $ \mathrm{rank}(\matt{P}) = \mathcal{R}$, the maximum number of distinct receive constellation points becomes $ \abs{\mathcal{Y}} = \mathcal{R} + 1 $. Thus, using a full rank precoder matrix, i.e., $ \matt{P} \in \mathbb{R}^{\Nt \times \Nt}  $, the receive constellation has $ \abs{\mathcal{Y}} =  \Nt +1 $ points and, in turn, we can achieve a rate closer to the capacity of the channel, $ C \leq \log_{2}(\Nt + 1) $ \gls{bpcu}, assuming a uniform distributed input signal.
      
This simple example motivated our study of whether higher-rank transmit covariance matrices can increase the system performance in the presence of the non-linear quantizers.

\subsection{Contributions}\label{sec:Intro_Sub_Contributions}
In this paper, we analyze and investigate the optimality of two aspects of traditional signal processing in 1-bit quantized \gls{MU}-\gls{MISO} systems: (i) proper signaling, and/or (ii) channel rank transmit covariance matrices.  We summarize our contributions presented in this paper for the downlink scenario as follows:
\begin{enumerate}
	\item In the first part of the paper, we investigate the structure of the transmit covariance matrix $ \matt{R}_{\bar{\vect{x}}} $.  We provide an achievable rate analysis, applying the Bussgang decomposition to investigate the sum rate lower-bound in 1-bit quantized \gls{MU}-\gls{MISO} systems. We investigate whether (i) improper signaling and/or (ii) higher-rank transmit covariance matrices maximize the sum rate lower-bound.  This analysis indicates that higher-rank transmit covariance matrices can improve the sum rate lower-bound, whereas improper signaling may only marginally improve the system performance.
	\item In the second part of the paper, we focus on the optimization of the linear precoder taking the results from our achievable rate analysis into account. In the end, we provide a gradient-projection algorithm to design a sub-optimal, higher-rank linear precoder.  To obtain a higher-rank linear precoder, we introduce the idea of a linear superposition matrix which allows for linear superposition coding in 1-bit quantized \gls{MU}-\gls{MISO} systems. Moreover, our higher-rank linear precoder shows the predicted performance increase due to the increase in rank, compared with the linear precoder \gls{TxWFQ} from \cite{Mezghani2009transmit}. 
\end{enumerate}
Our results indicate that indeed there are benefits in reconsidering signal processing methods for 1-bit quantized \gls{MU}-\gls{MISO} scenarios.  Moreover, we provide an algorithm to design a linear precoder \gls{TxWFQ}--$ \matt{\Pi} $ whose rank is higher than the rank of the channel.


\subsection{Paper Structure}\label{sec:Intro_Sub_Paper_Structure}
The rest of the paper is structured as follows.  In Section \ref{sec:System_Model} the system model is introduced, including the necessary mathematical tools required for our analysis.  In Section \ref{sec:Rate_Analysis} we delve into the achievable rate analysis, and in Section \ref{sec:transmit_Signal_Processing} we introduce our novel transmit precoder design taking the results from Section \ref{sec:Rate_Analysis} into account.  At the end of Section \ref{sec:transmit_Signal_Processing}, we show the performance improvements introduced by our linear precoder design, investigate the complexity of our algorithm, and analyze the robustness of the algorithm against channel estimation errors. Finally, in Section \ref{sec:Conclusions} we conclude the paper by summarizing our main results and providing an outlook onto further work.


\subsection{Notation}\label{sec:Intro_Sub_Notation}
Scalars, vectors and matrices are denoted by italic letters, bold italic lowercase letters and bold italic uppercase letters, respectively.  The operators $ (\cdot)^{\T}$, $ \tr{\cdot} $, $ \E[\cdot] $ $ \Re \lbrace \cdot \rbrace $, $ \Im \lbrace \cdot \rbrace $ represent the transpose, trace, expected value, real part and imaginary part, respectively. The notation $ \diag{\matt{A}} $ represents a diagonal matrix with the diagonal elements of $ \matt{A} $, while $ \nondiag{\matt{A}} $ represents the matrix $ \matt{A} - \diag{\matt{A}} $. The matrix operation $ \matt{A}^{\circ n} $ defines the Hadamard product to the $ n $th power, i.e., $ \matt{A}\circ \ldots \circ \matt{A} $, where $ [\matt{A}^{\circ n}]_{i,j} = a_{i,j}^{n} $, which represents element-wise multiplication.  The Kronecker product of two matrices is represented by $ \matt{A}\otimes\matt{B} $.
We use $ \matt{I}_{N} $ and $ \matt{0}_{N} $ to represent an $ N \times N  $ identity matrix and all-zero matrix, respectively.

Moreover, we introduce \gls{WL} notation (see e.g., \cite{Adali2011,mandic2009complex,hellings2015block}) to accommodate our analysis of whether proper signaling is optimal in quantized \gls{MU}-\gls{MISO} systems.  To this end, we introduce the following definitions:
\begin{definition}[Widely-Linear Vector]\label{def:WL_Vector}
	Taking a complex vector $ \vect{a} \in \mathbb{C}^{N} $, we can express it in \gls{WL} notation as
	\begin{align}\label{def:Widely_Linear_Vector}
	\bar{\vect{a}} 
	= \begin{bmatrix}
	\Re\lbrace\vect{a}\rbrace\\
	\Im\lbrace\vect{a}\rbrace
	\end{bmatrix}\in \mathbb{R}^{2 N}.
	\end{align}
\end{definition}

\begin{definition}[Strictly Linear Transformation]\label{def:WL_Matrix}
	A transformation in the complex domain is strictly linear, i.e., $ {{\vect{c}}= {\matt{B}}{\vect{a}} \in \mathbb{C}^{M} \Leftrightarrow \bar{\vect{c}}= \bar{\matt{B}}\bar{\vect{a}} \in \mathbb{R}^{2M} }$, \gls{iff}, in the real domain, the matrix $ \bar{\matt{B}} $ has the following structure
	\begin{align} \label{def:Widely_Linear_Matrix}
	\bar{\matt{B}} = \begin{bmatrix}
	\Re\lbrace\matt{B}\rbrace & -\Im\lbrace\matt{B}\rbrace \\
	\Im\lbrace\matt{B}\rbrace & \Re\lbrace\matt{B}\rbrace
	\end{bmatrix} 
	\in \mathbb{R}^{2 M \times 2  N}.
	\end{align}
\end{definition}

We define the real-valued covariance matrix of the arbitrary signal $ \bar{\vect{a}} $ in \gls{WL} notation as
\begin{align}\label{def:Covariance_Matrix}
\matt{R}_{\bar{\vect{a}}} = \begin{bmatrix}
\E[\Re \lbrace \vect{a} \rbrace \Re \lbrace \vect{a}^{\T} \rbrace ] & \E[\Re \lbrace \vect{a} \rbrace \Im \lbrace \vect{a}^{\T} \rbrace ]\\
\E[\Im \lbrace \vect{a} \rbrace \Re \lbrace \vect{a}^{\T} \rbrace ] & 
\E[\Im \lbrace \vect{a} \rbrace \Im \lbrace \vect{a}^{\T} \rbrace ]
\end{bmatrix}.
\end{align}
\begin{definition}[Proper Signals]\label{def:Proper}
	The signal $ {\vect{a}} $ is  proper \gls{iff} both of the following conditions hold:
\begin{align}
	\E\left [\Re \left \lbrace \vect{a} \right \rbrace 
	\Re \left \lbrace \vect{a}^{\T} \right\rbrace \right ] &= 
	\E\left[\Im \left \lbrace \vect{a} \right \rbrace 
	\Im \left \lbrace \vect{a}^{\T}\right  \rbrace \right ],\label{def:Proper_1} \\
	\E\left [\Re \left \lbrace \vect{a} \right \rbrace 
	\Im \left \lbrace \vect{a}^{\T} \right \rbrace \right ] &= 
	-\E\left [\Im \left \lbrace \vect{a} \right \rbrace 
	\Re \left \lbrace \vect{a}^{\T} \right \rbrace \right ].\label{def:Proper_2}
	\end{align}
\end{definition}


\section{System Model}\label{sec:System_Model}

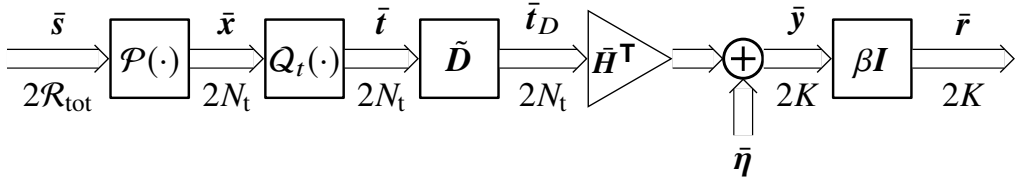
\begin{figure*}[!]
	\centering
	\resizebox{0.8\textwidth}{!}{
		\begin{tikzpicture}

	\usetikzlibrary[arrows.meta]
 	\tikzstyle{vecArrow} = [decoration={markings,
  	mark=at position 1 with {\arrow[] {Straight Barb[width=8pt,length=4pt]}}},
	double distance=5pt, shorten >= 3.08pt,
	preaction = {decorate},
	postaction = {draw,line width=5pt, white, shorten >= 3.08pt
	}]
	
    \tikzstyle{innerWhite} = [semithick, white,line width=1.4pt, shorten >= 4.5pt]
    
    \tikzstyle{triangle}=[
    isosceles triangle,  
    align=center,
    draw, 
    inner sep=1, 
    font=\small\sffamily\bfseries,
    isosceles triangle apex angle=60,
    isosceles triangle stretches]
    right iso/.style={ triangle,scale=0.5,sharp corners, anchor=center, xshift=-4mm};

	\def\stdx{3em}
	\def\stdy{-3em}
	
	\node[] (s)	at (0,0)	{};
	
	\node[dspsquare](Precoder) at ($ (s) + (1.5*\stdx,0) $) {$ \pcal(\cdot) $};
	\node[dspsquare](Qt)		at ($(Precoder) + (1.5*\stdx,0) $)	{$\qcal_{t}(\cdot)$};
	\node[dspsquare](D) at ($ (Qt) + (1.5*\stdx,0) $) {$ {\tilde{\matt{D}}} $};
	
	\node[triangle](H)		at ($ (D) + (1.5*\stdx,0) $)	{$ \bar{\matt{H}}^{\T} $};
	
	\node[dspadder](add) at ($ (H) + (1.25*\stdx,0) $){};
	\node[](noise) at ($ (add) + (0,\stdy) $) {$ \bar{\vect{\eta}} $};
	
	\node[dspsquare](beta)		at ($(add) + (1.25*\stdx,0) $)	{$\beta\matt{I}$};
	
	\node[](y)		at ($(beta) + (1.5*\stdx,0) $)	{};
	
	\draw[vecArrow] (s.east) -- (Precoder.west)node[below=0.35em,midway] {$2 \mathcal{R}_{\mathrm{tot}}$}node[above=0.35em,midway] {${\bar{\vect{s}}}$};
	\draw[vecArrow] (Precoder.east) -- (Qt.west)node[below=0.35em,midway] {$2N_{\mathrm{t}}$}node[above=0.35em,midway] {$\bar{\vect{x}}$};
	\draw[vecArrow] (Qt.east) -- (D.west)node[below=0.35em,midway] {$2\Nt$}node[above=0.35em,midway] {${\bar{\vect{t}}}$};
	\draw[vecArrow] (D.east) -- (H.west)node[below=0.35em,midway] {$2\Nt$}node[above=0.35em,midway] {${\bar{\vect{t}}_{D}}$};
	\draw[vecArrow] (H.east) -- (add.west) ;
	\draw[vecArrow] (noise.north) -- (add.south);
	\draw[vecArrow] ($ (add.east) + (0.01*\stdx,0) $) -- (beta.west) node[above=0.35em,midway] {$\bar{\vect{y}}$}node[below=0.35em,midway] {$2K$};
	\draw[vecArrow] (beta.east) -- (y.west) node[below=0.35em,midway] {$2K$}node[above=0.35em,midway] {$\bar{\vect{r}}$};

\end{tikzpicture}
	}
	\caption{Abstract Downlink Quantized \gls{MU}-\gls{MISO} System Model}
	\label{fig:MIMO_System_Model_TxMMSE}
\end{figure*}

We consider the downlink scenario of a single-cell, coarsely quantized \gls{MU}-\gls{MISO} system as depicted in Fig. \ref{fig:MIMO_System_Model_TxMMSE}.  The \gls{BS} has $ \Nt $ transmit antennas, each equipped with two 1-bit quantized \glspl{DAC} for the in-phase and quadrature signal components.  The \gls{BS} serves $ K $ single-antenna users simultaneously, and we assume the \glspl{ADC} at the users have infinite quantization resolution.  Furthermore, we assume that the \gls{BS} and the users are fully synchronized, with their \glspl{DAC} and \glspl{ADC} working at the same sampling frequency. 

Thus, assuming narrowband channels, we can collect the real-valued baseband received signals at each user into a single vector representation
\begin{align}\label{eq:y_WL}
\bar{\vect{y}} 
&= \bar{\matt{H}}^{\T}\tilde{\matt{D}}\bar{\vect{t}} + \bar{\vect{\eta}} 
\in \mathbb{R}^{2 K}.
\end{align}

The vector $ \bar{\vect{y}} \in \mathbb{R}^{2K}$ contains the received signals of all users, where $ [\bar{{y}}_{k}, \bar{{y}}_{k+K}]^{\T} = [\Re\lbrace{y}_{k}\rbrace, \Im\lbrace{y}_{k}\rbrace]^{\T} \in \mathbb{R}^{2} $ represents the received signal of user $ k $ in \gls{WL} notation.  The strictly linear (see Def. \ref{def:WL_Matrix}) downlink channel matrix is denoted by $ \bar{\vect{H}}^{\T} \in \mathbb{R}^{2 K \times 2\Nt} $ in \gls{WL} notation.  We assume perfect \gls{CSI} at the \gls{BS}\footnote{The impact of imperfect \gls{CSI} will be studied later in the numerical results.}, i.e., the matrix $ \bar{\matt{H}}^{\T} $ is perfectly known.  Furthermore, we assume that the complex channel elements are circular symmetric \gls{iid} Gaussian random variables with $ h_{k,n} = [\matt{H}]_{k,n} \sim \ccal\ncal_{\mathbb{C}} (0, 1), \forall k, n $.  The quantized transmit signal is $ {\bar{\vect{t}} \in \lbrace \pm1\rbrace^{2\Nt} }$, where we assume the output of the uniform 1-bit \glspl{DAC} is either $ \pm1 $.  

The diagonal, real-valued power allocation matrix is denoted by $ \tilde{\matt{D}} \in \mathbb{R}^{2\Nt \times 2\Nt} $.  We assume that $ \tilde{\matt{D}} $ does not necessarily have the strictly linear structure defined in Def. \ref{def:WL_Matrix}, so that the power can be allocated freely between the real and imaginary parts, which allows for improper signaling.  Improper signaling can also be achieved by introducing correlation between the real and imaginary parts of the signals before the \glspl{DAC}, i.e., breaking the circular symmetry of the complex signals.  Despite the fact that the power allocation matrix must be updated for every channel realization, one could still achieve constant envelope modulation per channel by feeding back a distinct scalar to the \glspl{PA} at each antenna, which adjusts the supply voltage at each \gls{PA} for a given channel, e.g.,  employing envelope tracking \glspl{PA} (see, e.g., \cite{raab2002power}).  These supply voltages remain constant during each channel coherence time.

Finally, the \gls{AWGN} is assumed to have the following distribution for all users: $ \bar{\vect{\eta}}  \sim \mathcal{C}\mathcal{N}(0, \sigma_{\eta}^{2}/2 \cdot \matt{I}_{2K}) $.  The scaling factor $ \beta \in \mathbb{R}_{+} $ at the receivers, introduced in \cite{joham2005linear}, can be interpreted as an automatic gain control which is required to amplify the received symbol at each user such that it lies the correct decision region.  We assume the scaling factor $ \beta $ changes relatively slowly and thus can be (perfectly) estimated over multiple received symbols by each user using blind methods \cite{amor201716,jacobsson2016nonlinear}, allowing the users to employ minimum distance decoding.

The real-valued input signal $ \bar{\vect{s}} \in \mathbb{R}^{2 \mathcal{R}_{\mathrm{tot}}} $ contains the $ 2 \mathcal{R}_{\mathrm{tot}} $ input symbols of all users.  We introduce $ \mathcal{R}_{\mathrm{tot}} = \sum_{k = 1}^{K} \mathcal{R}_{k}$ as the sum of the number of streams per user. The variable $ \mathcal{R}_{k} $ represents the number of streams each user receives, which, in turn, determines the increase in rank of the precoder matrix.   It should be noted that if $ \mathcal{R}_{k} = 1 $, the precoder matrix will have the same rank as the channel.  Moreover, $ \mathcal{R}_{k} \leq \Nt $ holds because the maximum number of streams per user is upper bounded by the number of transmit antennas.  Unless otherwise stated, the input signal is assumed to be Gaussian distributed with $ \bar{\vect{s}} \sim \ncal (\vect{0}, \matt{R}_{\bar{\vect{s}}} ) $.

The transmit signal $ \bar{\vect{x}} \in \mathbb{R}^{2\Nt} $ is the output of the precoder with input $ \bar{\vect{s}} $, i.e., $ \bar{\vect{x}} = \pcal(\bar{\vect{s}}), $ where $ \pcal(\cdot) $ is bijective but otherwise arbitrary and can be linear or non-linear. If we assume it to be a linear function, we do not restrict it to be strictly linear in the complex domain as defined in Def. \ref{def:WL_Matrix}, i.e., $ \tilde{\matt{P}} \neq \bar{\matt{P}} $.  This allows $ \matt{R}_{\bar{\vect{x}}} $ to have the arbitrary structure as in \eqref{def:Covariance_Matrix}.  

We define the non-linear quantization function in the 1-bit case to take the sign of the input signal $ \text{sign}(\bar{\vect{x}}) $, i.e.,
\begin{align}\label{eq:Def_Quantisation}
\mathcal{Q}_{t} :  \mathbb{R}^{2\Nt} \to \lbrace \pm 1 \rbrace^{2\Nt},\quad \bar{\vect{x}} \mapsto \mathcal{Q}_{t}(\bar{\vect{x}}) = \text{sign}(\bar{\vect{x}}) = \bar{\vect{t}}.
\end{align}
where the non-linear function $ \text{sign}(\cdot) $  is applied element-wise.  Therefore, the total power across all transmit antennas after quantization is $ \sum_{i = 1}^{2\Nt} \E[\abs{\bar{t}_{i}}^{2}] = 2\Nt$.  If we define the total available transmit power as $ E_{\mathrm{Tx}} $, and assume the power is equally allocated over all antennas, then the power allocation matrix must be a scaled identity matrix, i.e.,
\begin{align}
\tilde{\matt{D}} = \sqrt{\dfrac{E_{\mathrm{Tx}}}{2\Nt}}\matt{I}_{2\Nt}.
\end{align}
With this power allocation matrix we allow for improper signaling by introducing correlation between the transmit signal at different antennas.
Thus, the total power after the power allocation matrix is equal to $ {\sum_{i = 1}^{2\Nt} \E[\abs{\bar{t}_{D,i}}^{2}] = E_{\mathrm{Tx}}}$.


\subsection{Bussgang Decomposition} \label{sec:Bussgang}

Similar to previous work, e.g.,\cite{Mezghani2012capacity}, \cite{saxena2016analysis}, we model the quantization function using the Bussgang decomposition  \cite{Bussgang1952}.  According to the Bussgang theorem, the cross-correlation between two Gaussian distributed input signals remains the same when one signal is subjected to non-linear distortion, except for a scaling factor.  This implies that a non-linear function with Gaussian inputs can be modeled by a linear transformation and the addition of some distortion which is uncorrelated with the inputs.  

The transmit signal becomes approximately Gaussian distributed as the number of users increases
due to the central limit theorem.  Therefore, we assume $ {\bar{\vect{x}}  \sim \ncal(\vect{0}, \matt{R}_{\bar{\vect{x}}})}$ when $ K $ is large enough.  Thus, using the Bussgang theorem the quantization function in \eqref{eq:Def_Quantisation} can be modeled as
\begin{align}
\bar{\vect{t}} = \mathcal{Q}_{t}(\bar{\vect{x}})
= \matt{A}\bar{\vect{x}} + \vect{q},
\end{align}
where the quantization error, $ \vect{q} $ is uncorrelated with the input signal $ \bar{\vect{x}} $.  From the latter criterion we see that 
\begin{align}\label{eq:MMSE_Estimate_Bussgang}
\E\left[{\vect{q}}\bar{\vect{x}}^{\T}\right] 
= \matt{0}_{2\Nt} 
\Rightarrow \matt{A} 
= \matt{R}_{\bar{t}\bar{x}}\matt{R}_{\bar{x}}^{-1}.
\end{align}
Thus, we observe that the matrix $ \matt{A} $ is simply a linear \gls{MMSE} estimate of the quantized signal $ \bar{\vect{t}} $ from the unquantized input signal $ \bar{\vect{x}} $.
Moreover, the Bussgang decomposition depends on the covariance matrix between the quantized signal $ \bar{\vect{t}} $ and the unquantized signal $ \bar{\vect{x}} $.  

Finally, we can express the covariance matrix of the quantization error as
\begin{align}\label{eq:R_q}
\matt{R}_{\vect{q}} 
+ \matt{A}\matt{R}_{\bar{x}}\matt{A}^{\T}  \stackrel{\eqref{eq:MMSE_Estimate_Bussgang}}{=} \matt{R}_{\bar{t}} - \matt{R}_{\bar{t}\bar{x}}\matt{R}_{\bar{x}}^{-1}\matt{R}_{\bar{x}\bar{t}}
.
\end{align}


\subsection{Price's Theorem -- Quantized Covariance Matrices} \label{sec:Price_Theorem}
To calculate the covariance matrices $ \matt{R}_{\bar{\vect{t}}} = \E \left[\bar{\vect{t}}\bar{\vect{t}}^{\T} \right] $ and $ \matt{R}_{\bar{t}\bar{x}} = \E \left[\bar{\vect{t}}\bar{\vect{x}}^{\T} \right] $ we apply Price's theorem \cite{Price1958}, (for more details see e.g., \cite[Sec. II]{roth2015covariance}).  To this end, the covariance matrix of the quantized output signal is, e.g., \cite[p. 307]{papoulis2002probability}, 
\begin{align}\label{eq:R_t}
\matt{R}_{\bar{\vect{t}}} = \dfrac{2}{\pi}
\arcsin\left(
\diag{\matt{R}_{\bar{\vect{x}}}}^{-1/2}
\matt{R}_{\bar{\vect{x}}}
\diag{\matt{R}_{\bar{\vect{x}}}}^{-1/2}
\right),
\end{align}
where the factor $ 2/\pi $ comes from the fixed quantization levels and the real-valued function $ \arcsin(\matt{A}) $ is defined element-wise on the matrix argument $ \matt{A} $.  The covariance matrix between the input and output of the 1-bit quantizer can equally be calculated by applying Price's theorem
\begin{align}\label{eq:R_tx}
\matt{R}_{\bar{\vect{t}}\bar{\vect{x}}} &= \sqrt{\dfrac{2}{\pi}} \diag{\matt{R}_{\bar{\vect{x}}}}^{-1/2}\matt{R}_{\bar{\vect{x}}}.
\end{align}
Moreover, due to the real-valued \gls{WL} notation, the following relationship holds: $ \matt{R}_{\bar{\vect{x}}\bar{\vect{t}}} = \matt{R}_{\bar{\vect{t}}\bar{\vect{x}}}^{\T}  $.


\section{Achievable Rate Analysis}\label{sec:Rate_Analysis}
In this section, we investigate the achievable rate in 1-bit quantized \gls{MU}-\gls{MISO} systems by looking at the structure of the transmit covariance matrices $ \matt{R}_{\bar{\vect{x}}_{k}} $.  For our achievable rate analysis, we assume that the total transmit power is constant $ E_{\mathrm{Tx}} = 2\Nt $, and the \gls{SNR} at the receiver is varied by changing the noise variance $ \sigma_{\eta}^{2} $.  With equal power allocation, we have $ \tilde{\matt{D}} = \matt{I}_{2\Nt} $.  Moreover, since we are investigating the mutual information and not the signal processing techniques in this section, we assume the receiver scaling factor to be one, $ \beta = 1 $.  We assume that the \gls{CSI} is perfectly known at the \gls{BS} and at the users.

Using the Bussgang decomposition defined in Section \ref{sec:Bussgang} and the covariance matrices defined in Section \ref{sec:Price_Theorem}, we can express the real-valued received signal at user $ k $ from \eqref{eq:y_WL} as
\begin{align}
\bar{\vect{y}}_{k} 
& = \bar{\matt{H}}_{k}^{\T}(\matt{A}\bar{\vect{x}} + \vect{q}) + \bar{\vect{\eta}}_{k} \nonumber \\
& = \bar{\matt{H}}_{\mathrm{eff},k}^{\T} \bar{\vect{x}} + \tilde{\vect{\eta}}_{k}
,
\end{align}
where $ \bar{\matt{H}}_{k}^{\T} $ represents the strictly linear channel matrix of user $ k $.  We recall that the received signal at each single-antenna user and the total transmit signal are expressed in \gls{WL} notation from Def. \ref{def:WL_Vector}.  

Moreover, we introduce the effective channel $ \bar{\matt{H}}_{\mathrm{eff},k}^{\T} = \bar{\matt{H}}_{k}^{\T} \matt{A} \stackrel{\eqref{eq:MMSE_Estimate_Bussgang}}{=}\bar{\matt{H}}_{k}^{\T}\matt{R}_{\bar{t}\bar{x}}\matt{R}_{\bar{x}}^{-1}, $
and the effective noise $ \tilde{\vect{\eta}}_{k} = \bar{\matt{H}}_{k}^{\T}\vect{q} + \bar{\vect{\eta}}_{k} $, which is no longer Gaussian due to the quantization error, with $  \matt{R}_{\bar{x}} $ and $\matt{R}_{\bar{t}\bar{x}}$ defined in \eqref{eq:R_t} and \eqref{eq:R_tx}, respectively.
Furthermore, we express the transmit signal $ \bar{\vect{x}} $ as the sum of the transmit signals intended for each user, and we only consider coding schemes where the transmit signals for each user are independent, i.e., 
\begin{align}
\bar{\vect{x}} = \sum_{k = 1}^{K}\bar{\vect{x}}_{k} 
\quad \Rightarrow \quad 	\matt{R}_{\bar{\vect{x}}} 
= \sum_{k = 1}^{K}\matt{R}_{\bar{\vect{x}}_{k}}
,
\end{align}

\noindent where $ \bar{\vect{x}}_{k} $ and $ \matt{R}_{\bar{\vect{x}}_{k}} $ represent the transmit signal and transmit covariance matrix intended for user $ k $ prior to the \glspl{DAC}, respectively.


\subsection{Sum Rate Lower-Bound}

Now, we aim to calculate the mutual information between the signal intended for user $k$ and the signal that user receives, i.e., $ {I(\bar{\vect{x}}_{k}; \bar{\vect{y}}_{k}) = h(\bar{\vect{y}}_{k}) - h(\bar{\vect{y}}_{k}\vert\bar{\vect{x}}_{k}) }$, with the continuous entropy function $ h(\cdot) $.  We assume perfect knowledge of the \gls{CSI} at the \gls{BS} and at the users, and no cooperation between the users.  The encoding at the \gls{BS} does not use the non-causally known interference of the user's signals, i.e., we do not employ dirty paper coding, and the users decode the received signal by treating the \gls{MUI} as noise.

We first focus on the second continuous entropy term 
\begin{align}\label{eq:Cond_Entropy}
h(\bar{\vect{y}}_{k}\vert\bar{\vect{x}}_{k}) & = h\left (\left.\bar{\matt{H}}_{\mathrm{eff},k}^{\T} \sum_{k = 1}^{K}\bar{\vect{x}}_{k} + \tilde{\vect{\eta}}_{k}\right \vert\bar{\vect{x}}_{k}\right ) \nonumber \\
& \stackrel{(a)}{\leq} h\left (\bar{\matt{H}}_{\mathrm{eff},k}^{\T} \sum_{\substack{l = 1\\l \neq k}}^{K}\bar{\vect{x}}_{l} + \tilde{\vect{\eta}}_{k}\right ) 
,
\end{align}
where inequality $ (a) $ comes from the fact that conditioning cannot increase entropy, \cite[Th. 2.6.5]{Cover2012}, and holds with equality if $ \tilde{\vect{\eta}}_{k} $ and $ \bar{\vect{x}}_{k} $ are statistically independent.  Moreover, the addition of a constant term does not change the entropy, i.e., $ h(\bar{\vect{x}}_{k}\vert\bar{\vect{x}}_{k}) = 0 $.  However, despite the fact that $ \vect{q} $ and $ \bar{\vect{x}}_{k} $ are uncorrelated, they may still be dependent. The total noise is $ \bar{\matt{H}}_{\mathrm{eff},k}^{\T} \sum_{\substack{l = 1, l \neq k}}^{K}\bar{\vect{x}}_{l} + \tilde{\vect{\eta}}_{k}$, which contains the \gls{MUI}, quantization error $ \vect{q} $, and the \gls{AWGN}.   

Furthermore, in \cite{Diggavi2001} (see also \cite{li2017downlink}), it was shown that for a given noise covariance matrix, Gaussian distributed noise minimizes the mutual information in a given system.   Thus, assuming Gaussian distributed inputs and total noise from \eqref{eq:Cond_Entropy},  we can write the instantaneous mutual information lower-bound of the Gaussian system as
\begin{align} \label{eq:MuI_LB}
I(\bar{\vect{x}}_{k}; \bar{\vect{y}}_{k}) \geq 
\dfrac{1}{2} 
\log_{2}\det(\matt{I}_{2} + \mathrm{SQINR}_{k}) 
,
\end{align}
where the \gls{SQINR}$ _{k} $ defined in \eqref{eq:SQINR_Def} (on the next page) shows the contribution of the \gls{MUI}, \gls{QE} and \gls{AWGN}.  The identity matrix $ \matt{I}_{2} $ comes from the fact that we consider the real and imaginary parts separately.
\begin{figure*}[!]
	\begin{align}\label{eq:SQINR_Def}
	\mathrm{{SQINR}}_{k}\! = 
	\left(\underbrace{\bar{\matt{H}}_{\mathrm{eff},k}^{\T}
	\sum_{\substack{l = 1,l \neq k}}^{K}\matt{R}_{\bar{\vect{x}}_{l}}\bar{\matt{H}}_{\mathrm{eff},k}}_{\mathrm{\gls{MUI}}} + \underbrace{\bar{\matt{H}}_{k}^{\T}\matt{R}_{\vect{q}}\bar{\matt{H}}_{k}}_{\mathrm{\gls{QE}}} + \underbrace{\matt{R}_{\bar{\vect{\eta}}_{k}}}_{\mathrm{\gls{AWGN}}}\right)^{-1}
	\bar{\matt{H}}_{\mathrm{eff},k}^{\T}\matt{R}_{\bar{\vect{x}}_{k}} \bar{\matt{H}}_{\mathrm{eff},k}
	\end{align}
\end{figure*}

With the mutual information lower-bound defined per user in \eqref{eq:MuI_LB}, we can now express the instantaneous sum rate lower-bound by summing over all $ k = 1, \ldots, K $:
\begin{align} \label{eq:Def_WSR}
\sum_{k = 1}^{K}
I(\bar{\vect{x}}_{k}; \bar{\vect{y}}_{k})
 \geq
  \dfrac{1}{2}\sum_{k = 1}^{K}
  \log_{2}\det(\matt{I}_{2} + \text{\gls{SQINR}}_{k})
  .
\end{align}

Finally, we optimize for the transmit covariance matrices which maximize the sum rate lower-bound from \eqref{eq:Def_WSR}
\begin{align}\label{eq:WSR_OP}
\matt{R}_{\bar{\vect{x}}_{k,\mathrm{opt}}} = 
\argmax_{\matt{R}_{\bar{\vect{x}}_{k}} 
\succeq \matt{0}, \forall k} \left \lbrace
\sum_{k = 1}^{K}
\log_{2}\det(\matt{I}_{2} + \text{\gls{SQINR}}_{k})
\right \rbrace
,
\end{align}

\noindent where the optimization is performed over all positive semi-definite transmit covariance matrices for all users, i.e., $ {\matt{R}_{\bar{\vect{x}}_{k}} \succeq \matt{0}}, \forall k $.  Since we assume perfect \gls{CSI} at the \gls{BS} and at the users, we calculate the ergodic achievable sum rate as the average of the maximum sum rates achieved per channel realization, i.e., we average the sum rate lower-bounds over the channel realizations \cite[Sec. II-C1]{goldsmith2003capacity}.

Since the argument of the $ \log_{2}\det(\cdot) $ function in the sum rate depends non-linearly on the user's transmit covariance matrix (see \eqref{eq:R_t}), we wish to investigate whether traditional signal processing techniques still maximize \eqref{eq:WSR_OP}, i.e., whether channel rank transmit covariance matrices and proper signaling are still optimal.


\subsection{Cholesky Decomposition}
First, we note that
\begin{align}
\rank{\matt{R}_{\bar{\vect{x}}}} = \rank{\sum_{k = 1}^{K}\matt{R}_{\bar{\vect{x}}_{k}}} \leq \sum_{k = 1}^{K}\rank{\matt{R}_{\bar{\vect{x}}_{k}}},
\end{align}
and we define the Cholesky decompositon of $ \matt{R}_{\bar{\vect{x}}_{k}} $ to be $ \matt{R}_{\bar{\vect{x}}_{k}} = \matt{L}_{k}(\mathcal{R}_{k})\matt{L}_{k}^{\T}(\mathcal{R}_{k}) $, 
where $ \mathcal{R}_{k} $ denotes the number of streams for user $ k $, which determines the rank of $ \matt{R}_{\bar{\vect{x}}_{k}} $. The rank of the transmit covariance matrix can be varied by observing the structure of the Cholesky factor: 
\begin{align}\label{eq:MU_MISO_Chol_L}
\matt{L}_{k}(\mathcal{R}_{k})  = \begin{bmatrix}
l_{1,1} & 0 & \ldots & 0 & \ldots & 0\\
l_{2,1} &  \ddots & \ddots & \vdots & \ddots & \vdots \\
l_{3,1} &  \ddots & l_{2\mathcal{R}_{k},2\mathcal{R}_{k}} & 0 &\ldots & 0\\
l_{4,1} &  \ddots & l_{2(\mathcal{R}_{k}+1),2\mathcal{R}_{k}} & 0  &\ldots & 0\\
\vdots & \ddots & \vdots & \vdots & \ddots & \vdots \\
l_{2\Nt,1} & \ldots & l_{2\Nt,2\mathcal{R}_{k}} & 0 & \ldots & 0
\end{bmatrix},
\end{align}
where $ l_{i,i} > 0 $ for $ i \leq \mathcal{R}_{k}\leq \Nt $ and $ l_{i,j} \in \real $ for $ i > j, j \leq \mathcal{R}_{k} \leq \Nt $.  
Finally, we can restate \eqref{eq:WSR_OP} as
\begin{align}\label{eq:WSR_OP_Chol}
\matt{L}_{k, \mathrm{opt}}(\mathcal{R}_{k}) = \argmax_{\matt{L}_{k}(\mathcal{R}_{k}), \forall k} \Bigg \lbrace
\dfrac{1}{2}
\sum_{k = 1}^{K}
\log_{2}\det(\matt{I}_{2} + \text{\gls{SQINR}}_{k})
\Bigg \rbrace,
\end{align}
where the $ \text{\gls{SQINR}}_{k} $ is now parameterized in terms of $ \matt{L}_{k}(\mathcal{R}_{k}) $, for a given $ \mathcal{R}_{k} $, instead of the transmit covariance matrix $ \matt{R}_{\bar{\vect{x}}_{k}} $.  When considering proper signaling, we can also restrict user $ k $ to employ proper signaling by further restricting the covariance matrix $ \matt{R}_{\bar{\vect{x}}_{k}} = \matt{L}_{k}(\mathcal{R}_{k})\matt{L}_{k}(\mathcal{R}_{k})^{\T} $ to fulfill \eqref{def:Proper_1} and \eqref{def:Proper_2} from Def. \ref{def:Proper}.


\subsection{Simulation Results: Achievable Rate}
In this section, we provide numerical results for our achievable rate analysis in 1-bit quantized \gls{MU}-\gls{MISO} systems.  We simulated a downlink scenario with a $ \Nt =16 $ antenna \gls{BS} and $ K = 2 $ single antenna users, solving the optimization in \eqref{eq:WSR_OP_Chol} numerically for both proper and improper transmit covariance matrices.  In this scenario, the rank of the user's transmit covariance matrices must be $ \mathcal{R}_{k} \leq \Nt =  16 $.  In our simulations we assumed the rank of each users' transmit covariance matrix was equal, i.e., $ \mathcal{R}_{1} = \mathcal{R}_{2} = \mathcal{R} $, and set $ \mathcal{R} = 1 $ and $ 2 $.  We observed that if we increased the rank beyond 2, the additional streams per user led to worse performance.  We plot the ergodic sum rate lower-bound by averaging the sum rate lower-bounds over $ 200 $ \gls{iid} channel realizations where we set $ E_{\mathrm{Tx}} = 2\Nt $ and varied the noise variance $ \sigma_{\eta}^{2} \in \lbrace -20, \ldots, 20 \rbrace $ dB.   

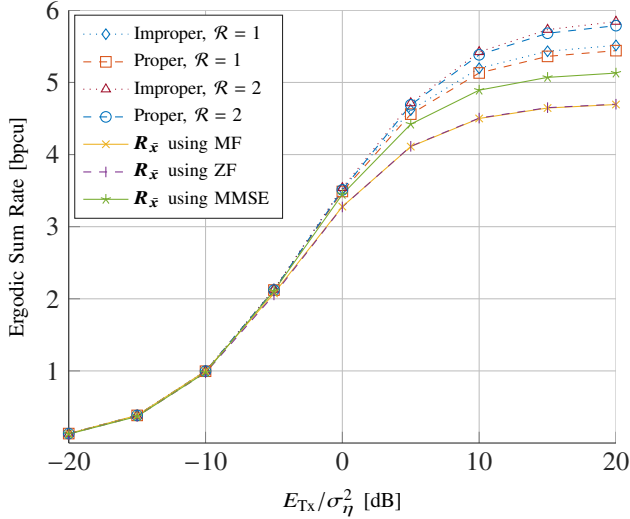
\begin{figure}
	\centering
%
%
%
\definecolor{mycolor1}{rgb}{0.00000,0.44700,0.74100}%
\definecolor{mycolor2}{rgb}{0.85000,0.32500,0.09800}%
\definecolor{mycolor3}{rgb}{0.92900,0.69400,0.12500}%
\definecolor{mycolor4}{rgb}{0.49400,0.18400,0.55600}%
\definecolor{mycolor5}{rgb}{0.46600,0.67400,0.18800}%
\definecolor{mycolor6}{rgb}{0.30100,0.74500,0.93300}%
\definecolor{mycolor7}{rgb}{0.63500,0.07800,0.18400}%
\tikzset{every mark/.append style={solid}}
\pgfplotsset{tick label style={font=\footnotesize}, label style={font=\footnotesize}, legend style={font=\scriptsize}}
\pgfplotsset{every axis legend/.append style={at={(0.01,0.99)},anchor=north west}}
\begin{tikzpicture}[every pin/.style={fill=white}]

\begin{axis}[%
width=\columnwidth,
height=0.3\textheight,
every outer x axis line/.append style={white!15!black},
every x tick label/.append style={font=\color{white!15!black}},
xmin=-20,
xmax=20,
xlabel={$ E_{\mathrm{Tx}}/\sigma_{\eta}^{2} $ [dB]},
xmajorgrids,
every outer y axis line/.append style={white!15!black},
every y tick label/.append style={font=\color{white!15!black}},
ymin=0,
ymax=6,
ytick={1,2,...,6},
ylabel={Ergodic Sum Rate [bpcu]},
ymajorgrids,
axis x line*=bottom,
axis y line*=left,
legend style={draw=white!15!black,fill=white,legend cell align=left}
]
\addplot [color=mycolor1,dotted,mark=diamond,mark options={solid}]
table[row sep=crcr]{%
	-20	0.133185063600043\\
	-15	0.385699457438236\\
	-10	0.999611374757037\\
	-5	2.13015826774963\\
	0	3.52218802429894\\
	5	4.61682813992181\\
	10	5.19706354602939\\
	15	5.43258913375736\\
	20	5.51382447555567\\
};
\addlegendentry{Improper, $ \mathcal{R} = 1 $};
\addplot [color=mycolor2,dashed,mark=square,mark options={solid}]
table[row sep=crcr]{%
	-20	0.131087572525572\\
	-15	0.383180834430734\\
	-10	0.995838345495712\\
	-5	2.12149366487506\\
	0	3.48863248085168\\
	5	4.5607256918317\\
	10	5.13158206551744\\
	15	5.36160028362833\\
	20	5.44112627997886\\
};
\addlegendentry{Proper, $ \mathcal{R} = 1 $};
\addplot [color=mycolor7,dotted,mark=triangle,mark options={solid}]
table[row sep=crcr]{%
	-20	0.128493351448774\\
	-15	0.380165147844658\\
	-10	0.994110496594735\\
	-5	2.12762709271173\\
	0	3.53725120463209\\
	5	4.71365244713954\\
	10	5.42331386490235\\
	15	5.73217745744093\\
	20	5.84363149435772\\
};
\addlegendentry{Improper, $ \mathcal{R} = 2 $};
\addplot [color=mycolor1,dashed,mark=o,mark options={solid}]
table[row sep=crcr]{%
	-20	0.130579158795776\\
	-15	0.38243980471978\\
	-10	0.99717875270412\\
	-5	2.12111999287377\\
	0	3.50968162758119\\
	5	4.68952177433654\\
	10	5.38334180799874\\
	15	5.68214124822071\\
	20	5.78744690653306\\
};
\addlegendentry{Proper, $ \mathcal{R} = 2 $};

\addplot [color=mycolor3,solid, mark=x]
table[row sep=crcr]{%
-20	0.131043512378674\\
-15	0.384115459535842\\
-10	0.996981876761721\\
-5	2.074322743512\\
0	3.27837497353805\\
5	4.11056979205141\\
10	4.50063226408803\\
15	4.64557486157298\\
20	4.69410861617905\\
};
\addlegendentry{$ \matt{R}_{\bar{\vect{x}}} $ using MF};

\addplot [color=mycolor4,dashed, mark=|]
table[row sep=crcr]{%
-20	0.125601440556061\\
-15	0.370553087710265\\
-10	0.97372825252282\\
-5	2.05432261686582\\
0	3.27356825901337\\
5	4.1144655371294\\
10	4.5057698713133\\
15	4.65035811306064\\
20	4.69864625606695\\
};
\addlegendentry{$ \matt{R}_{\bar{\vect{x}}} $ using ZF};

\addplot [color=mycolor5,solid, mark=star]
table[row sep=crcr]{%
-20	0.124298972279208\\
-15	0.368396154753645\\
-10	0.979075000228218\\
-5	2.1091284143605\\
0	3.44828302883148\\
5	4.42196731569053\\
10	4.89290821978497\\
15	5.07034139461093\\
20	5.13004757594618\\
};
\addlegendentry{$ \matt{R}_{\bar{\vect{x}}} $ using MMSE};

\end{axis}
\end{tikzpicture}%
	\caption{\gls{MU}-\gls{MISO} Downlink lower-bound of the ergodic sum rate with $ \Nt = 16 $ and $ K = 2 $, averaged over $ 200 $ \gls{iid} channels.}
	\label{fig:MU_MISO_Chol_16x2}
\end{figure}

We plot the performance of different transmit covariance matrices in \figref{fig:MU_MISO_Chol_16x2}, comparing our optimized covariance matrices, $ \matt{R}_{\bar{\vect{x}}_{k,\mathrm{opt}}} = \matt{L}_{k, \mathrm{opt}} \matt{L}^{\T}_{k, \mathrm{opt}}$,  with the traditional strictly linear \gls{MF}, \gls{ZF} and \gls{MMSE} precoders.  We observe that at low-\gls{SNR} our optimized transmit covariance matrices converge to those employing traditional signal processing techniques.  This indicates that traditional signal processing methods, i.e., channel rank transmit covariance matrices and proper signaling, may be optimal at low-\gls{SNR} in the \gls{MU} scenario, similar to the \gls{SU} scenario (see \cite[Th. 1]{decandido2017are}).  
 
At mid- to high-\gls{SNR} we observe that the optimized covariance matrices diverge from the traditional signal processing techniques.  The transmit covariance matrix using the \gls{MMSE} precoder shows a higher sum rate lower-bound at high-\gls{SNR} compared with the other two traditional signal processing techniques, but our optimized covariance matrices provide better performance.  The gain at higher-\gls{SNR} for the improper and proper solutions with channel rank, i.e., $ \mathcal{R} = 1 $, is due to the fact that our solutions further mitigate the \gls{MUI}.

Furthermore, we observe that when the rank of both users' transmit covariance matrices is higher than the rank of the channel, i.e., $ \mathcal{R} = 2 $, the sum rate lower-bound is the highest.  This indicates that higher-rank transmit covariance matrices can achieve better performance in 1-bit quantized \gls{MU}-\gls{MISO} scenarios.  These results concur with our results in the \gls{SU}-\gls{MISO} scenario in \cite[Sec. V]{decandido2017are}.  Moreover, improper signaling only seems to marginally improve the performance.  

We can summarize the results as follows: (i) higher-rank transmit covariance matrices maximize the sum rate lower-bound, and (ii) improper signaling only marginally improves the sum rate lower-bound.  Thus, in the following we will attempt to optimize a linear precoder matrix which has a rank higher than the rank of the channel, and further investigate whether improper signaling can improve the uncoded-\gls{BER} or \gls{MSE} performance.


\section{Transmit Signal Processing}\label{sec:transmit_Signal_Processing}

In this section, we move on from our investigation of the transmit covariance matrix and introduce a linear precoder design taking the results from Section \ref{sec:Rate_Analysis} into account.  Therefore, we assume that the precoder function is a linear precoder matrix $ \tilde{\matt{P}} \in \mathbb{R}^{2\Nt \times 2 \mathcal{R}_{\mathrm{tot}}}$.  Note that the \gls{WL} precoder matrix can have an arbitrary structure and not the \gls{SL} structure defined in Def. \ref{def:WL_Matrix}, i.e., $ \tilde{\matt{P}} \neq \bar{\matt{P}} $.  To this end, we express the received signal as
\begin{align}\label{eq:MU_MISO_received_r}
\bar{\vect{r}} = \beta \bar{\vect{y}} = \beta\left(
\bar{\matt{H}}^{\T}\tilde{\matt{D}}\mathcal{Q}_{t}(\tilde{\matt{P}}\bar{\vect{s}}) + \bar{\vect{\eta}}
\right) \in \mathbb{R}^{2K}.
\end{align}


\subsection{Superposition Matrix}
With the definition of the received signal in \eqref{eq:MU_MISO_received_r} we can define the \gls{MSE}, $ \varepsilon $, as
\begin{align}
\varepsilon = \E\left[ \Vert \bar{\vect{r}} - \matt{\Pi}\bar{\vect{s}} \Vert^{2}_{2}\right],
\end{align}
where we introduce the linear superposition matrix $ \matt{\Pi} $ which allows the users to receive symbols from higher-order constellations than those transmitted (see e.g., \cite{amor201716,huan2009cooperative,sun2012superposition,wang2008optimizing}). 

In our transmitter signal processing design we take the results from \cite[Fig. 2(a)]{jacobsson2016nonlinear} into account which show that with a linear precoder (\gls{ZF}) and 1-bit \glspl{DAC} at the \gls{BS}, \gls{QPSK} transmit symbols show the best uncoded-\gls{BER} performance. Therefore, we assume that the input signal in our system are \gls{QPSK} for all users.

Assuming all users receive the same constellation, i.e., $ \mathcal{R}_{k} = \mathcal{R} \; \forall k$, the linear superposition matrix describing higher order \gls{QAM} based on \gls{QPSK} is defined as
\begin{align} \label{eq:Tx_Pi}
\matt{\Pi} =  \matt{I}_{2 K}\otimes \vect{\tau}^{\T} \in \mathbb{R}^{2 K \times 2 K \mathcal{R}}
\end{align}
and the superposition row vector $ \vect{\tau}^{\T} $ is defined as
\begin{align}\label{eq:Def_Superposition vector}
\vect{\tau}^{\T} = \begin{bmatrix}
2^{\mathcal{R}-1}& 2^{\mathcal{R}-2}& \ldots&  
 2^{1}& 2^{0}
\end{bmatrix}
\in \mathbb{R}^{1\times \mathcal{R}}
.
\end{align}
The superposition vector $ \vect{\tau}^{\T}\in \lbrace 1, 2,4, \ldots, 2^{\mathcal{R}-1}\rbrace^{1\times \mathcal{R}} $ has a length $ \mathcal{R} $ per user which determines the rank of each user's precoder.  The maximum rank of the precoder is equal to the number of transmit antennas, i.e., $ \mathcal{R} \leq \Nt $.

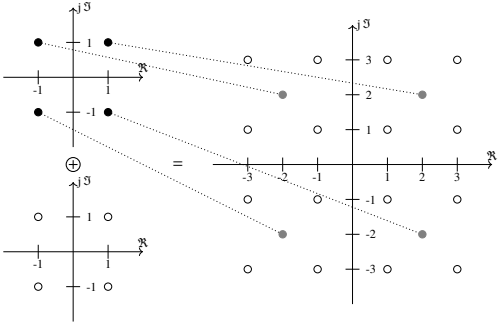
\begin{figure}
	\centering
	\resizebox{0.8\columnwidth}{!}{
		\begin{tikzpicture}[thick]

\def\stdx{3};
\def\stdy{2.5};

\def\stdxx{5};

\draw[->] ($ (-2,0) + (-\stdx,\stdy) $) -- ($ (2,0) + (-\stdx,\stdy) $) node[above] {$ \Re $};
\draw[->] ($ (0,-2) + (-\stdx,\stdy) $) -- ($ (0,2) + (-\stdx,\stdy) $) node[right] {$ \jim\Im $};

\foreach \x in {-1,1}
\draw ($ (\x,-0.20) + (-\stdx,\stdy) $) -- ($ (\x,0.20) + (-\stdx,\stdy) $) node[below=0.3] {\x};

\foreach \y in {-1,1}
\draw ($ (-0.20,\y) + (-\stdx,\stdy) $) -- ($ (0.20,\y) + (-\stdx,\stdy) $) node[right=0.05] {\y};

\foreach \x in {-1,1}
\foreach \y in {-1,1}
{
	\filldraw ($ (\x,\y) + (-\stdx,\stdy) $) circle (3pt);
}

\draw[->] ($ (-2,0) + (-\stdx,-\stdy) $) -- ($ (2,0) + (-\stdx,-\stdy) $) node[above] {$ \Re $};
\draw[->] ($ (0,-2) + (-\stdx,-\stdy) $) -- ($ (0,2) + (-\stdx,-\stdy) $) node[right] {$ \jim\Im $};

\foreach \x in {-1,1}
\draw ($ (\x,-0.20) + (-\stdx,-\stdy) $) -- ($ (\x,0.20) + (-\stdx,-\stdy) $) node[below=0.3] {\x};

\foreach \y in {-1,1}
\draw ($ (-0.20,\y) + (-\stdx,-\stdy) $) -- ($ (0.20,\y) + (-\stdx,-\stdy) $) node[right=0.05] {\y};

\foreach \x in {-1,1}
\foreach \y in {-1,1}
{
	\draw ($ (\x,\y) + (-\stdx,-\stdy) $) circle (3pt);
}

\foreach \x in {-1,1}
\foreach \y in {-1,1}
{
	\draw[dotted] ($ (\x,\y) + (-\stdx,\stdy) $) --  ($ (2*\x,2*\y) + (\stdxx,0) $);
}

\node at ($ (-2,0) + (-\stdx,\stdy) +(-0.5,0) $) {};
\node at ($ (0,0) $) {\Large $ = $};
\draw[->] ($ (-4,0) + (\stdxx,0) $) -- ($ (4,0) + (\stdxx,0) $) node[above] {$ \Re $};
\draw[->] ($ (0,-4) + (\stdxx,0) $) -- ($ (0,4) + (\stdxx,0) $) node[right] {$ \jim\Im $};

\foreach \x in {-3,-2,-1,1,2,3}
\draw ($ (\x,-0.20) + (\stdxx,0) $) -- ($ (\x,0.20) + (\stdxx,0) $) node[below=0.3] {\x};

\foreach \y in {-3,-2,-1,1,2,3}
\draw ($ (-0.20,\y) + (\stdxx,0) $) -- ($ (0.20,\y) + (\stdxx,0) $) node[right=0.05] {\y};

\foreach \x in {-3,-1,1,3}
\foreach \y in {-3,-1,1,3}
{
	\draw ($ (\x,\y) + (\stdxx,0) $) circle (3pt);
}
\node[dspadder](add) at($ (0,-2) + (-\stdx, 2) $) {};

\foreach \x in {-2,2}
\foreach \y in {-2,2}
{
	\filldraw[gray] ($ (\x,\y) + (\stdxx,0) $)  circle (3pt);
}

\end{tikzpicture}
	}
	\caption{Linear Superposition Matrix -- Two \gls{QPSK} symbols to one 16--\gls{QAM} symbol}
	\label{fig:Superposition}
\end{figure}

To clarify how our linear superposition matrix works, assume $ \mathcal{R} = 2 $ which implies that each user receives 16-\gls{QAM} symbols, and we have $ \vect{\tau}^{\T} = [2, 1]$ as per \eqref{eq:Def_Superposition vector}.  We observe in \figref{fig:Superposition} how the superposition vector $ \vect{\tau}^{\T} = [2, 1] $ works; (i) the first symbol (solid points) is multiplied by a factor 2 and defines which quadrant the received symbol should lie in, (ii) the second symbol (hollow points) is added to the first and defines which 16-\gls{QAM} symbol should be received.  Since we assume \gls{QPSK} input symbols and the specific superposition matrix defined in \eqref{eq:Tx_Pi}, the superimposed received symbols will be $ M $-\gls{QAM}, where $ M $ depends on the chosen number of streams per user, $ \mathcal{R}_{k} $.

\subsection{MSE Definition}
With the superposition matrix defined in \eqref{eq:Tx_Pi}, we can express the \gls{MSE} as
\begin{align}\label{eq:Tx_MSE_Final}
\varepsilon 
=&\beta^{2}\dfrac{2}{\pi}\tr{
	\bar{\matt{H}}^{\T}
	\tilde{\matt{D}}
	\arcsin(\matt{P}'\matt{P}'^{\T})
	\tilde{\matt{D}}
	\bar{\matt{H}}}  + \beta^{2}\tr{\matt{R}_{\bar{\vect{\eta}}}} + 
\tr{\matt{\Pi}\matt{R}_{\bar{\vect{s}}}\matt{\Pi}^{\T}}\nonumber 
\\
&
- 2\beta\sqrt{\dfrac{2}{\pi}}
\tr{
	\bar{\matt{H}}^{\T}
	\tilde{\matt{D}}
	\matt{P}'\matt{R}_{\bar{\vect{s}}}^{1/2}\matt{\Pi}^{\T}
},
\end{align}
where we have inserted the covariance matrices of the quantized signals defined in \eqref{eq:R_t} and \eqref{eq:R_tx}; also we define the normalized precoding matrix $ \matt{P}' $ as
\begin{align} \label{eq:TxWFQ_P'}
\matt{P}' =\diag{\tilde{\matt{P}}\matt{R}_{\bar{\vect{s}}}\tilde{\matt{P}}^{\T}}^{-1/2}\tilde{\matt{P}}\matt{R}_{\bar{\vect{s}}}^{1/2}.
\end{align}


\subsection{Transmit Wiener Filter Design}
In this subsection we introduce our algorithm to calculate a higher-rank version of the \gls{TxWFQ} from \cite{Mezghani2009transmit}.  First, we define the optimization problem as follows
\begin{align}
\lbrace \tilde{\matt{P}}_{\mathrm{opt}},\beta_{\mathrm{opt}}, \tilde{\matt{D}}_{\mathrm{opt}}\rbrace =
\argmin_{\tilde{\matt{P}}, \beta, \tilde{\matt{D}}} \; \lbrace \varepsilon\rbrace \; 
\;\mathrm{s.t.}\;
 \begin{array}{l}
 \E\left [\left 
\lVert \bar{\vect{t}}_{D}
\right \rVert_{2}^{2}
\right ] \leq E_{\mathrm{Tx}},\\
\tilde{\matt{D}} \in \mathbb{R}^{2\Nt \times 2\Nt} \text{ is diagonal,}
 \end{array}
\end{align}
with $ \varepsilon $ defined in \eqref{eq:Tx_MSE_Final} and the sum power constraint is applied after the power allocation matrix $ \tilde{\matt{D}} $ (see \figref{fig:MIMO_System_Model_TxMMSE}).

Intuitively, we understand that the power allocated to the transmit antennas by the precoder is normalized back to unit power by the 1-bit \glspl{DAC}.  Therefore, we choose $ \tilde{\matt{D}} $ to restore the desired power allocation of the precoder by setting
\begin{align}\label{eq:TxWFQ_D_opt}
\tilde{\matt{D}}_{\mathrm{opt}} = \diag{\tilde{\matt{P}}\matt{R}_{\bar{\vect{s}}}\tilde{\matt{P}}^{\T}}^{1/2}.
\end{align}
With the choice of the power allocation matrix $ \tilde{\matt{D}}_{\mathrm{opt}} $ defined in \eqref{eq:TxWFQ_D_opt}, we can rewrite the optimization problem as
\begin{align} \label{eq:TxWFQ_OP_final}
\lbrace \tilde{\matt{P}}_{\mathrm{opt}},\beta_{\mathrm{opt}}\rbrace =
\argmin_{\tilde{\matt{P}}, \beta} \; \lbrace \varepsilon\rbrace \; 
\st
\begin{array}{l}
\tr{\tilde{\matt{P}}\matt{R}_{\bar{\vect{s}}}\tilde{\matt{P}}^{\T}} \leq E_{\mathrm{Tx}},\\
\tilde{\matt{D}}_{\mathrm{opt}} = \diag{\tilde{\matt{P}}\matt{R}_{\bar{\vect{s}}}\tilde{\matt{P}}^{\T}}^{1/2},
\end{array}
\end{align}
where the sum-power constraint comes from the fact that ${ \bar{\vect{t}}_{D} = \tilde{\matt{D}}_{\mathrm{opt}}\bar{\vect{t}} }$ and using the optimal power allocation matrix from \eqref{eq:TxWFQ_D_opt}.


\subsection{Arcsine Approximation}\label{sec:TxSP_Arcsine_Approx}
We note that due to the non-linear matrix function $ \arcsin(\cdot) $ in the \gls{MSE} expression \eqref{eq:Tx_MSE_Final}, the derivative of $ \varepsilon $ \gls{wrt} $ \tilde{\matt{P}} $ proves difficult to solve for in closed form.  Therefore, we use the second-order Taylor expansion of the off-diagonal elements defined as: $ \arcsin(x) \approx x + {1}/{6}\cdot x^{3} $.  The diagonal elements are given by: $ \arcsin\left (\diag{\matt{P}'\matt{P}'^{\T}}\right ) = \arcsin(\matt{I}_{2\Nt}) = {\pi}/{2}\cdot\matt{I}_{2\Nt} $.  Thus, the matrix $ \arcsin(\cdot) $ function can be approximated as
\begin{align}\label{eq:TxWFQ_Arcsin_Approx}
\arcsin(\matt{P}'\matt{P}'^{\T})  \approx \matt{P}'\matt{P}'^{\T} +
\dfrac{1}{6}\left(\matt{P}'\matt{P}'^{\T}\right)^{\circ 3}
+
\left(\dfrac{\pi}{2} - \dfrac{7}{6}\right)\matt{I}_{2\Nt}
,
\end{align}
where $ \matt{A}^{\circ n} $ represents the matrix Hadamard product to the power $ n $.  We use the second-order Taylor expansion since we want to retain the non-linearities introduced by the coarse quantization to observe the performance gains from higher-rank transmit covariance matrices.
Therefore, we can substitute the optimal power allocation matrix from \eqref{eq:TxWFQ_D_opt} and the approximation from \eqref{eq:TxWFQ_Arcsin_Approx} into the \gls{MSE} expression from \eqref{eq:Tx_MSE_Final} to arrive at
\begin{align}\label{eq:Tx_MSE_Final_approx}
\varepsilon \approx &
\beta^{2}\dfrac{2}{\pi}
\tr{
	\bar{\matt{H}}^{\T}\left (
	\tilde{\matt{P}}\matt{R}_{\bar{\vect{s}}}\tilde{\matt{P}}^{\T} + \left (\dfrac{\pi}{2}-\dfrac{7}{6}\right ) \diag{\tilde{\matt{P}}\matt{R}_{\bar{\vect{s}}}\tilde{\matt{P}}^{\T}}
	\right )
	\bar{\matt{H}}}
\nonumber \\
&+
\beta^{2}\dfrac{2}{\pi}\dfrac{1}{6}
\tr{
	\bar{\matt{H}}^{\T}\left (
	\tilde{\matt{D}}_{\mathrm{opt}}^{-2}\left(
	\tilde{\matt{P}}\matt{R}_{\bar{\vect{s}}}\tilde{\matt{P}}^{\T}
	\right)^{\circ 3}\tilde{\matt{D}}_{\mathrm{opt}}^{-2}
	\right )
	\bar{\matt{H}}}
  \nonumber \\
& 
- 2\beta\sqrt{\dfrac{2}{\pi}}
\tr{\bar{\matt{H}}^{\T}\tilde{\matt{P}}\matt{R}_{\bar{\vect{s}}}\matt{\Pi}^{\T}}
+ \beta^{2}\tr{\matt{R}_{\bar{\vect{\eta}}}} + 
\tr{\matt{\Pi}\matt{R}_{\bar{\vect{s}}}\matt{\Pi}^{\T}}.
\end{align}


\subsection{Gradient Projection Algorithm}
Since the optimization problem in \eqref{eq:TxWFQ_OP_final} is non-convex and non-linear \gls{wrt} the precoder matrix $ \tilde{\matt{P}} $, solving \eqref{eq:TxWFQ_OP_final} in closed-form is intractable.  Consequently, we use a gradient-projection algorithm \cite[p. 466]{boyd2004convex} to iteratively solve for a locally optimal solution, with a projection back onto the feasible set.  The gradient-projection algorithm we implement is outlined in Algorithm~\ref{alg:gradient_Proj_WFQHR}.

\begin{algorithm}
	\begin{algorithmic}[1]
		\State \textbf{Initialization:}
		\State $ \tilde{\matt{P}}_{(0)} $, $ \beta_{(0)} \gets g^{*}(\tilde{\matt{P}}_{(0)})$, $ \gamma = 10 $ and $ n = 0 $
		\Repeat
		\If{$ \varepsilon_{(n+1)} \leq \varepsilon_{(n)} $} \label{alg:Check_MSE}
		\State $ \tilde{\matt{P}}_{(n+1)} \gets \mathcal{P}_{C}\left (\tilde{\matt{P}}_{(n)} - \gamma\frac{\partial\varepsilon\left (\tilde{\matt{P}}_{(n)}, \beta^{*}_{(n)}\right )}{\partial \tilde{\matt{P}}} \right ) $  \label{alg:Step_Projection_FR1}
		\State $ \beta^{*}_{(n+1)} \gets 
		g^{*}\left (\tilde{\matt{P}}_{(n)}\right )
		$ \Comment{defined in Appendix \ref{sec:Appendix}}\label{alg:Update_beta_FR1}
		\State $ n \gets n + 1 $
		\Else
		\State $ \gamma \gets \gamma/2$ \label{alg:Backtrack_gamma}
		\EndIf
		\Until{$ {\vert \varepsilon_{(n+1)} - \varepsilon_{(n)} \vert}/{\varepsilon_{(n)}} \leq \delta$ }
	\end{algorithmic}
	\caption{Gradient Projection Algorithm to Solve for the Higher-Rank, \gls{WL} \gls{TxWFQ}--$ \matt{\Pi} $}
	\label{alg:gradient_Proj_WFQHR}
\end{algorithm}

First, we initialize our algorithm with a random full rank matrix $ \tilde{\matt{P}}_{(0)} $ and calculate the initial scaling factor $ \beta_{(0)} $ using the function $ g^{*}(\tilde{\matt{P}}) $.  This function finds the optimal scaling factor for a given precoder matrix and will be introduced in Appendix \ref{sec:Appendix} (see \eqref{eq:TxWFQHR_g*}).  We use an initial constant step-size of $ \gamma = 10 $, but we allow for backtracking in Step \ref{alg:Backtrack_gamma}, where we halve the gradient step-size if the \gls{MSE} in iteration $ (n+1) $ is larger than in the current iteration $ (n) $, which is checked in Step \ref{alg:Check_MSE}. 

In Step \ref{alg:Step_Projection_FR1} we update the precoder by taking a step in the direction of the \gls{MSE} gradient \gls{wrt} the precoder, where the derivative term is defined in Appendix \ref{sec:Appendix}.  Here, the projection function $ \mathcal{P}_{C}(\cdot) $ ensures that the sum-power constraint $ \tr{\tilde{\matt{P}}\matt{R}_{\bar{\vect{s}}}\tilde{\matt{P}}^{\T}} \leq E_{\mathrm{Tx}} $ is fulfilled in each iteration.  The optimal scaling factor is updated in Step \ref{alg:Update_beta_FR1} using the function $ g^{*}(\tilde{\matt{P}}) $ defined in Appendix \ref{sec:Appendix}.  Our algorithm runs until the stopping criterion is met, which is triggered once the relative difference in \gls{MSE} from the previous iteration is less than a predefined threshold, $ \delta $.


\subsection{Simulation Results: Signal Processing}
In this subsection, we present simulation results for the \gls{TxWFQ}--$ \matt{\Pi} $ precoder introduced above.  We compare our linear precoder design with the \gls{TxWFQ} design from \cite{Mezghani2009transmit}. We note the following facts about the precoder design from \cite{Mezghani2009transmit}: (i) it has channel rank, (ii) it is strictly linear, and (iii) the authors further optimize the quantization output step-sizes, i.e., the outputs of the \glspl{DAC} are not uniform.  In the end, the resulting power allocation in \cite{Mezghani2009transmit} is equivalent to the optimal power allocation matrix $ \tilde{\matt{D}}_{\mathrm{opt}} $ we introduced in  \eqref{eq:TxWFQ_D_opt}.  Additionally, we plot the optimal \gls{TxWF} introduced in \cite{joham2005linear}, simulating the \gls{TxWF} in an unquantized scenario, i.e., assuming the \glspl{DAC} at the \gls{BS} have infinite quantization resolution, which we will refer to as \gls{TxWF} (unq.). 

We assume that the \gls{BS} has $ \Nt = 128 $ transmit antennas, which serves $ K = 4 $ single antenna users. Our simulation results assume a constant noise covariance matrix $ \matt{R}_{\vect{\eta}} =\matt{I}_{K} $, and vary the transmit power $ E_{\mathrm{Tx}} \in \lbrace 0, \ldots, 21 \rbrace$ dB. In our simulations, we used a block length of $ N_{\mathrm{b}} = 10,000 $ symbols and averaged over 200 \gls{iid} channel realizations. We terminate our algorithms with the value $ \delta = 10^{-4} $.

We plot two solutions for our precoder design, a \gls{WL} and a \gls{SL} solution.  The \gls{SL} solution has the same structure as defined in Def. \ref{def:WL_Matrix}.  To obtain a \gls{SL} solution we use the fact that the \gls{SL} structure in Def. \ref{def:WL_Matrix} is maintained under multiplication, addition, transposition and inversion as shown in \cite{hellings2015block} and \cite{hellings2015iterative}.  Therefore, if we initialize  our algorithm with a \gls{SL} matrix, i.e., $ \tilde{\matt{P}}_{(0)} = \bar{\matt{P}}_{(0)} $ with the structure from Def. \ref{def:WL_Matrix}, then the resulting solution will be \gls{SL}.

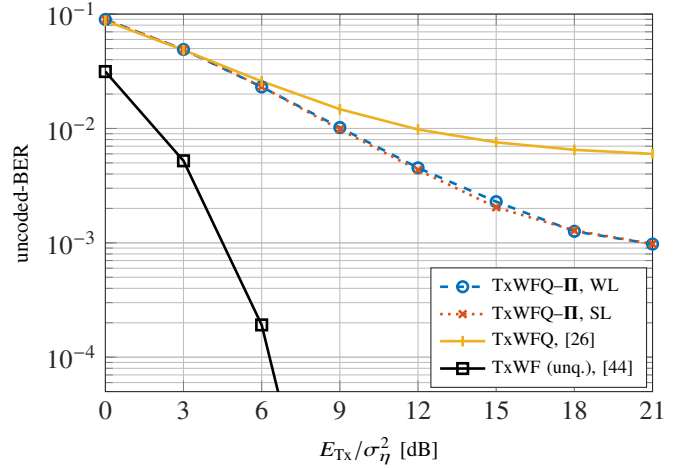
\begin{figure}
	\centering
%
%
%
\definecolor{mycolor1}{rgb}{0.00000,0.44700,0.74100}%
\definecolor{mycolor2}{rgb}{0.85000,0.32500,0.09800}%
\definecolor{mycolor3}{rgb}{0.92900,0.69400,0.12500}%
\definecolor{mycolor4}{rgb}{0.49400,0.18400,0.55600}%
\tikzset{every mark/.append style={solid}}
\pgfplotsset{tick label style={font=\footnotesize}, label style={font=\footnotesize}, legend style={font=\scriptsize}}
\pgfplotsset{every axis legend/.append style={at={(0.99,0.01)},anchor=south east}}
\begin{tikzpicture}

\begin{axis}[%
width=\columnwidth,
height=0.27\textheight,
every outer x axis line/.append style={white!15!black},
every x tick label/.append style={font=\color{white!15!black}},
xmin=0,
xmax=21,
xtick={0,3,...,21},
xlabel={$ E_{\mathrm{Tx}}/\sigma_{\eta}^{2} $ [dB]},
xmajorgrids,
every outer y axis line/.append style={white!15!black},
every y tick label/.append style={font=\color{white!15!black}},
ymode=log,
ymin=5e-05,
ymax=0.1,
yminorticks=true,
ylabel={uncoded-{BER}},
ymajorgrids,
yminorgrids,
title style={font=\bfseries},
legend style={draw=white!15!black,fill=white,legend cell align=left}
]
\addplot [color=mycolor1,dashed,line width=1.0pt,mark=o,mark options={solid}]
  table[row sep=crcr]{%
0	0.08974715625\\
3	0.049012375\\
6	0.023082\\
9	0.010192\\
12	0.0045239375\\
15	0.00229084375\\
18	0.00126121875\\
21	0.000975312500000001\\
};
\addlegendentry{\gls{TxWFQ}--$ \matt{\Pi} $, \gls{WL}};
\addplot [color=mycolor2,dotted,line width=1.0pt,mark=x,mark options={solid}]
  table[row sep=crcr]{%
0	0.089753625\\
3	0.04913784375\\
6	0.023231375\\
9	0.00980909375\\
12	0.0043365625\\
15	0.00204509375\\
18	0.001283375\\
21	0.00097803125\\
};
\addlegendentry{\gls{TxWFQ}--$ \matt{\Pi} $, \gls{SL}};
\addplot [color=mycolor3,solid,line width=1.0pt,mark=|]
  table[row sep=crcr]{%
0	0.08746784375\\
3	0.04850134375\\
6	0.02583028125\\
9	0.014746375\\
12	0.00979578124999999\\
15	0.00756878125\\
18	0.00650603125\\
21	0.00599478125\\
};
\addlegendentry{\gls{TxWFQ}, \cite{Mezghani2009transmit}};
\addplot [color=black,solid,line width=1.0pt,mark=square]
  table[row sep=crcr]{%
0	0.0314460625\\
3	0.00521740625\\
6	0.000191625\\
9	3.125e-07\\
12	0\\
15	0\\
18	0\\
21	0\\
};
\addlegendentry{\gls{TxWF} (unq.), \cite{joham2005linear}};
\end{axis}
\end{tikzpicture}%
	\caption{\gls{MU}-\gls{MISO} Downlink 16-QAM uncoded-BER using Alg.~\ref{alg:gradient_Proj_WFQHR} with $ \Nt = 128 $ and $ K = 4 $, averaged over $ 200 $ \gls{iid} channels.}
	\label{fig:MU_MISO_TxWFQ_128x4_BER}
\end{figure}

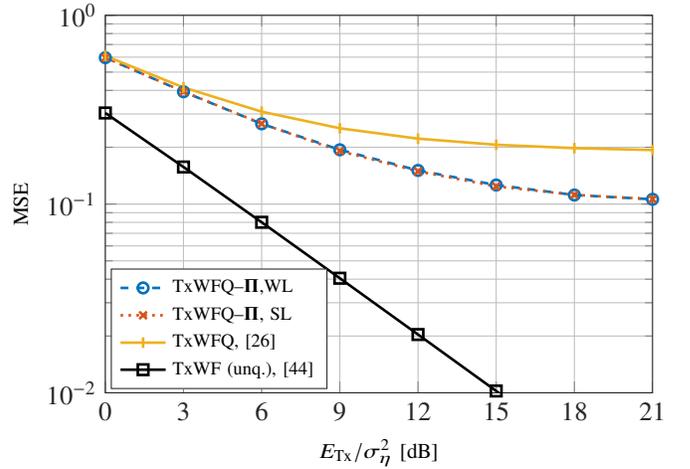
\begin{figure}
	\centering
%
%
%
\definecolor{mycolor1}{rgb}{0.00000,0.44700,0.74100}%
\definecolor{mycolor2}{rgb}{0.85000,0.32500,0.09800}%
\definecolor{mycolor3}{rgb}{0.92900,0.69400,0.12500}%
\definecolor{mycolor4}{rgb}{0.49400,0.18400,0.55600}%
\tikzset{every mark/.append style={solid}}
\pgfplotsset{tick label style={font=\footnotesize}, label style={font=\footnotesize}, legend style={font=\scriptsize}}
\pgfplotsset{every axis legend/.append style={at={(0.01,0.01)},anchor=south west}}
\begin{tikzpicture}

\begin{axis}[%
width=\columnwidth,
height=0.27\textheight,
every outer x axis line/.append style={white!15!black},
every x tick label/.append style={font=\color{white!15!black}},
xmin=0,
xmax=21,
xtick={0,3,...,21},
xlabel={$ E_{\mathrm{Tx}}/\sigma_{\eta}^{2} $ [dB]},
xmajorgrids,
every outer y axis line/.append style={white!15!black},
every y tick label/.append style={font=\color{white!15!black}},
ymode=log,
ymin=0.01,
ymax=1,
yminorticks=true,
ylabel={\gls{MSE}},
ymajorgrids,
yminorgrids,
title style={font=\bfseries},
legend style={draw=white!15!black,fill=white,legend cell align=left}
]
\addplot [color=mycolor1,dashed,line width=1.0pt,mark=o,mark options={solid}]
  table[row sep=crcr]{%
0	0.596338179996316\\
3	0.393637075895063\\
6	0.266100715716847\\
9	0.193867055383924\\
12	0.150704149671934\\
15	0.126142517056168\\
18	0.111654108830035\\
21	0.105973921578077\\
};
\addlegendentry{\gls{TxWFQ}--$ \matt{\Pi} $,\gls{WL}};
\addplot [color=mycolor2,dotted,line width=1.0pt,mark=x,mark options={solid}]
  table[row sep=crcr]{%
0	0.596469610863819\\
3	0.394255016881568\\
6	0.266727404853432\\
9	0.19083082123965\\
12	0.148976409210744\\
15	0.124073986174077\\
18	0.111873742492189\\
21	0.106171404281502\\
};
\addlegendentry{\gls{TxWFQ}--$ \matt{\Pi} $, \gls{SL}};
\addplot [color=mycolor3,solid,line width=1.0pt,mark=|]
  table[row sep=crcr]{%
0	0.610093659842942\\
3	0.414698495166752\\
6	0.308555854659415\\
9	0.252010710783501\\
12	0.222019854501411\\
15	0.206096087806821\\
18	0.197702198739428\\
21	0.193314053929025\\
};
\addlegendentry{\gls{TxWFQ}, \cite{Mezghani2009transmit}};
\addplot [color=black,solid,line width=1.0pt,mark=square]
  table[row sep=crcr]{%
0	0.303760783349785\\
3	0.157153051927617\\
6	0.0800600417002444\\
9	0.0404593806115611\\
12	0.020362843179556\\
15	0.0102271365205418\\
18	0.00513114533468707\\
21	0.00257303445375036\\
};
\addlegendentry{\gls{TxWF} (unq.), \cite{joham2005linear}};
\end{axis}
\end{tikzpicture}%
	\caption{\gls{MU}-\gls{MISO} Downlink 16-QAM \gls{MSE} using Alg.~\ref{alg:gradient_Proj_WFQHR} with $ \Nt = 128 $ and $ K = 4 $, averaged over $ 200 $ \gls{iid} channels.}
	\label{fig:MU_MISO_TxWFQ_128x4_MSE}
\end{figure}

First, we present uncoded-\gls{BER} and \gls{MSE} results using 16-\gls{QAM} symbols in \figref{fig:MU_MISO_TxWFQ_128x4_BER} and \figref{fig:MU_MISO_TxWFQ_128x4_MSE}, respectively.  To receive 16-\gls{QAM} symbols, the \gls{BS} transmits two \gls{QPSK} symbols to each user using the superposition vector from \eqref{eq:Def_Superposition vector} as $ \vect{\tau}^{\T} = [2, 1] $, implying $ \mathcal{R} = 2 $.  Thus, the rank of each user's precoder matrix is twice the rank of the channel.  In \figref{fig:MU_MISO_TxWFQ_128x4_BER}, we observe that our precoder design from Alg.~\ref{alg:gradient_Proj_WFQHR} outperforms the linear \gls{TxWFQ} method at higher \gls{SNR}.  We observe that this increase in uncoded-\gls{BER} and \gls{MSE} performance, compared with the \gls{TxWFQ} design from \cite{Mezghani2009transmit}, is due to the increase in rank of our \gls{TxWFQ}--$ \matt{\Pi} $ precoder design. However, there appears to be no gain from using improper signaling, i.e., the \gls{WL} solution performed as well as the \gls{SL} solution.  

Interestingly, both the \gls{WL} and the \gls{SL} solutions turn out to be independent of the random initialization.  Our solutions achieve an uncoded-\gls{BER} of $ 10^{-2} $ at around 9 dB, which is roughly 3 dB better than \gls{TxWFQ}.  Compared with the unquantized \gls{TxWF} we see roughly a 7 dB performance loss due to the quantization.  In \figref{fig:MU_MISO_TxWFQ_128x4_MSE}, we also observe similar performance gains in terms of \gls{MSE} for our higher-rank precoder design over the whole \gls{SNR} range.


\subsubsection{Complexity Analysis}
\begin{table}
	\centering
	\begin{tabular}{| l ||   c  | c | c  |c  | c | c | c | } 
		\hline 	
		$ E_{\mathrm{{Tx}}}/\sigma_{\eta}^{2} $ [dB]	 & 3 & 6 &9& 12 &15& 18   &21  \\
		\hline \hline
		Alg.~\ref{alg:gradient_Proj_WFQHR} \gls{WL}  &  33  &  86  & 111 &  137  & 166  & 209  & 254 \\
		\hline
		Alg.~\ref{alg:gradient_Proj_WFQHR} \gls{SL}  &   30  &  91  & 117  & 138  & 170  & 201  & 248\\
		\hline
	\end{tabular}
	\caption{Alg.~\ref{alg:gradient_Proj_WFQHR} with $ \Nt = 128 $ and $ K = 4 $ with 16-\gls{QAM} Average Number of Iterations}
	\label{tab:128x4_16QAM_Iters}
\end{table}
In Table \ref{tab:128x4_16QAM_Iters}, we show the average number of iterations, including the back-tracking steps, for  Alg.~\ref{alg:gradient_Proj_WFQHR} using either a \gls{WL} or \gls{SL} initialization.  We observe that the number of iterations grows almost linearly with the \gls{SNR} and at high-\gls{SNR} roughly 250 iterations are required to achieve a relative \gls{MSE} difference of $ \delta = 10^{-4} $.  Moreover, we see that, on average, both the \gls{WL} and \gls{SL} solutions require roughly the same number of iterations to converge.

Moreover, we take a closer look at the computational complexity of Alg.~\ref{alg:gradient_Proj_WFQHR} and calculate an asymptotic upper bound on the number of \gls{flops} required.  We observe that most of the computational complexity comes from calculating the derivative of the \gls{MSE} \gls{wrt} the precoder matrix (derived in Appendix \ref{sec:Appendix}).   Due to the term in \eqref{eq:db_dP}, we see that the asymptotic upper bound is: \[ \mathcal{O}\left (\Nt^{2}\cdot K \cdot \mathcal{R}_{\mathrm{tot}}\right ) \text{ \gls{flops}}, \] 
which is quadratic in the number of antennas but linear in the number of users and total number of streams per user.  It should be noted that our derivation of the derivative of the \gls{MSE} \gls{wrt} the precoder was not optimized to consider the number of \gls{flops} required, and there may be more efficient implementations.


\subsubsection{Channel State Estimation Error}
Thus far, we have assumed perfect \gls{CSI} at the \gls{BS}.  In the following, we investigate how sensitive our algorithm is to \gls{CSI} estimation errors.  To this end, we introduce the estimated channel matrix $ \bar{\matt{H}}_{\mathrm{est}} $ as
\begin{align}
\bar{\matt{H}}_{\mathrm{est}} = \sqrt{1 - \xi}\bar{\matt{H}} + \sqrt{\xi}\bar{\matt{\Gamma}},
\end{align}
where $ \xi \in [ 0, 1 ] $ and $ \left [\bar{\matt{\Gamma}}\right ]_{i,j} \sim \mathcal{C}\mathcal{N}(0,1), \forall i,j $.  The variable $ \xi $ represents the variance of the channel estimation error, where a value $ \xi = 0 $ is equivalent to a system without estimation error, i.e., perfect \gls{CSI}, and $ \xi = 1 $ is a fully erroneous channel estimation, i.e., where the \gls{BS} has no \gls{CSI}.  Intermediate values of $ \xi $ represent partial \gls{CSI} estimation errors.

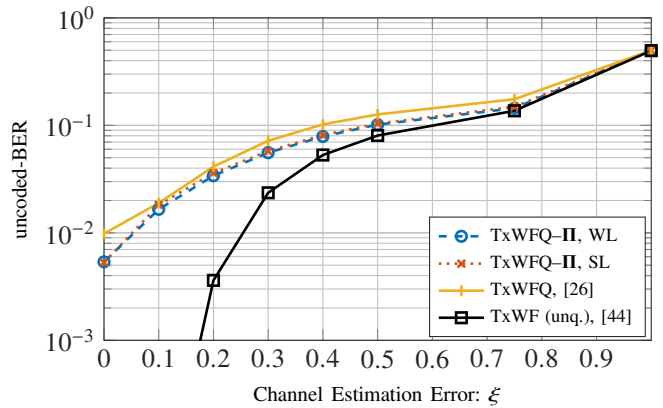
\begin{figure}
	\centering
%
%
%
\definecolor{mycolor1}{rgb}{0.00000,0.44700,0.74100}%
\definecolor{mycolor2}{rgb}{0.85000,0.32500,0.09800}%
\definecolor{mycolor3}{rgb}{0.92900,0.69400,0.12500}%
\definecolor{mycolor4}{rgb}{0.49400,0.18400,0.55600}%
\tikzset{every mark/.append style={solid}}
\pgfplotsset{tick label style={font=\footnotesize}, label style={font=\footnotesize}, legend style={font=\scriptsize}}
\pgfplotsset{every axis legend/.append style={at={(0.99,0.01)},anchor=south east}}
\begin{tikzpicture}

\begin{axis}[%
width=\columnwidth,
height=0.24\textheight,
every outer x axis line/.append style={white!15!black},
every x tick label/.append style={font=\color{white!15!black}},
xmin=0,
xmax=1,
xtick={0,0.1,...,1},
xlabel={Channel Estimation Error: $ \xi $},
xmajorgrids,
every outer y axis line/.append style={white!15!black},
every y tick label/.append style={font=\color{white!15!black}},
ymode=log,
ymin=1e-03,
ymax=1,
yminorticks=true,
ylabel={uncoded-BER},
ymajorgrids,
yminorgrids,
title style={font=\bfseries},
legend style={draw=white!15!black,fill=white,legend cell align=left}
]
\addplot [color=mycolor1,dashed,line width=1.0pt,mark=o,mark options={solid}]
table[row sep=crcr]{%
0	0.00536825\\
0.1	0.01654978125\\
0.2	0.0339929375\\
0.3	0.0556705625\\
0.4	0.0788175\\
0.5	0.10161178125\\
0.75	0.144566125\\
1	0.49861590625\\
};
\addlegendentry{\gls{TxWFQ}--$ \matt{\Pi} $, \gls{WL}};
\addplot [color=mycolor2,dotted,line width=1.0pt,mark=x,mark options={solid}]
table[row sep=crcr]{%
0	0.00528921875\\
0.1	0.0181863125\\
0.2	0.036010875\\
0.3	0.05763996875\\
0.4	0.08121753125\\
0.5	0.1038819375\\
0.75	0.14922540625\\
1	0.49820959375\\
};
\addlegendentry{\gls{TxWFQ}--$ \matt{\Pi} $, \gls{SL}};
\addplot [color=mycolor3,solid,line width=1.0pt,mark=|]
table[row sep=crcr]{%
0	0.0098251875\\
0.1	0.018952375\\
0.2	0.0412898125\\
0.3	0.07195740625\\
0.4	0.10212503125\\
0.5	0.12637746875\\
0.75	0.17520678125\\
1	0.49864325\\
};
\addlegendentry{\gls{TxWFQ}, \cite{Mezghani2009transmit}};
\addplot [color=black,solid,line width=1.0pt,mark=square]
table[row sep=crcr]{%
0	0\\
0.1	1.753125e-05\\
0.2	0.003610125\\
0.3	0.02358071875\\
0.4	0.05297209375\\
0.5	0.0804468125\\
0.75	0.136631875\\
1	0.4956566875\\
};
\addlegendentry{\gls{TxWF} (unq.), \cite{joham2005linear}};
\end{axis}
\end{tikzpicture}%
	\caption{\gls{MU}-\gls{MISO} Downlink uncoded-\gls{BER} vs. Channel Estimation Error employing 16-\gls{QAM} modulation at $ E_{\mathrm{Tx}}/\sigma_{\eta}^{2} = 12 $ dB.}
	\label{fig:MU_MISO_TxWFQ_128x4_BER_CSI}
\end{figure}

We plot the sensitivity of our algorithm against \gls{CSI} estimation error in \figref{fig:MU_MISO_TxWFQ_128x4_BER_CSI} for 16-\gls{QAM} received symbols at $ E_{\mathrm{Tx}}/\sigma_{\eta}^{2} = 12 $ dB.  We observe that our algorithm shows a slightly better performance compared with the \gls{TxWFQ} solution for all $ \xi $ values, although larger performance gains are seen with smaller \gls{CSI} estimation errors, since an increase in \gls{CSI} estimation error can also be seen as a decrease in \gls{SNR}.

\section{Conclusion}\label{sec:Conclusions}
In this paper, we reconsider linear transmit signal processing methods in 1-bit quantized \gls{MU}-\gls{MISO} downlink scenarios using an achievable rate analysis.  Our results indicate that higher-rank precoders can increase the lower-bound of the achievable sum rate.  By taking these results into account, we developed an algorithm to design a higher-rank linear precoder.  The derived precoder achieved performance superior to a state-of-the-art linear signal processing method with channel rank, both in terms of uncoded-\gls{BER} and \gls{MSE}.  These gains were due to the higher rank of the linear precoders; we observed no additional gain by employing improper signaling.  For 16-\gls{QAM} symbols, we observe a 3 dB gain for an uncoded-\gls{BER} of $ 10^{-2} $ over traditional linear signal processing techniques for a system with $ \Nt = 128 $ \gls{BS} antennas and $ K = 4 $ single antenna users.

Non-linear precoding methods where the transmit vector is optimized symbol-by-symbol, (e.g., \cite{jedda2016minimum,jedda2017massive,jacobsson2016quantized,casteneda20171-bit,tirkkonen2017subset,landau2017branch,amor201716,jacobsson2016nonlinear}), show uncoded-\gls{BER} and \gls{MSE} performance even closer to the unquantized \gls{TxWF}. However, they require much higher computational complexity as they must work at the sampling rate and scale with the number of transmit antennas.  In comparison, the linear precoder design presented here motivates reconsidering traditional linear precoder designs to improve the system performance with low complexity.  Moreover, the linear precoder matrix only has to be calculated once per channel coherence time instead of for each input symbol, which drastically reduces the computational complexity.

To extend the work presented in this paper, one could analyze a system using higher resolution \glspl{DAC}, still assuming constant envelope modulation.  Additionally, one could optimize the superposition matrix $ \matt{\Pi} $. In the end, $ \matt{\Pi} $ determines the increase in rank of the precoder matrix, and also depends on the users' channels which could be taken into account during the optimization.  Finally, it would be interesting to extend the work to frequency selective channels employing \gls{OFDM}; initial results show that \gls{OFDM} can be implemented with low-resolution \glspl{DAC} and linear processing (e.g., \cite{guerreiro2016use,jacobsson2017massive}).


\appendix[Derivations in Algorithm 1]\label{sec:Appendix}

In this Appendix we briefly derive the various functions and derivatives required in Alg.~\ref{alg:gradient_Proj_WFQHR}.
Note that, in the following derivations we will drop the iteration index $ (n) $ for notational brevity.  First, we restate the \gls{MSE} term from \eqref{eq:Tx_MSE_Final} as
\begin{align}\label{eq:Appendix_MSE}
\varepsilon = \beta^{2} \left (a(\tilde{\matt{P}}) + b(\tilde{\matt{P}}) + d\right ) - 2\beta c(\tilde{\matt{P}}) + e,
\end{align}
where we define the following functions
\begin{align}
a(\tilde{\matt{P}}) &\coloneq \dfrac{2}{\pi}
\tr{
	\bar{\matt{H}}^{\T}\left (
	\tilde{\matt{P}}\matt{R}_{\bar{\vect{s}}}\tilde{\matt{P}}^{\T} + \left (\dfrac{\pi}{2}-\dfrac{7}{6}\right ) \diag{\tilde{\matt{P}}\matt{R}_{\bar{\vect{s}}}\tilde{\matt{P}}^{\T}}
	\right )
	\bar{\matt{H}}} \label{eq:appendix_a}\\
b(\tilde{\matt{P}}) & \coloneq
\dfrac{2}{\pi}\dfrac{1}{6}
\tr{
	\bar{\matt{H}}^{\T}\left (
	\tilde{\matt{D}}_{\mathrm{opt}}^{-2}\left(
	\tilde{\matt{P}}\matt{R}_{\bar{\vect{s}}}\tilde{\matt{P}}^{\T}
	\right)^{\circ 3}\tilde{\matt{D}}_{\mathrm{opt}}^{-2}
	\right )
	\bar{\matt{H}}}
\label{eq:appendix_b}
 \\
c(\tilde{\matt{P}}) & \coloneq
-\sqrt{\dfrac{2}{\pi}}
\tr{\bar{\matt{H}}^{\T}\tilde{\matt{P}}\matt{R}_{\bar{\vect{s}}}\matt{\Pi}^{\T}}
\label{eq:appendix_c}
\\
d &= \tr{\matt{R}_{\bar{\vect{\eta}}}}  \text{ and } 
e =
\tr{\matt{\Pi}\matt{R}_{\bar{\vect{s}}}\matt{\Pi}^{\T}}.\label{eq:appendix_e}
\end{align}
We note that the functions $  a(\tilde{\matt{P}}) $ and $  b(\tilde{\matt{P}}) $ come from the second-order Taylor expansion of the non-linear $ \arcsin(\cdot) $ function (see Section \ref{sec:TxSP_Arcsine_Approx}).  Moreover, we recall that the optimal power allocation matrix $ \tilde{\matt{D}}_{\mathrm{opt}} $ is also a function of the precoder matrix.

With the approximate \gls{MSE} expression from \eqref{eq:Appendix_MSE}, we can define the function $ g^{*}(\tilde{\matt{P}}) $ by setting $ \partial \varepsilon / \partial \beta =0$, yielding
\begin{align}\label{eq:TxWFQHR_g*}
g^{*}\left (\tilde{\matt{P}}\right ) \coloneqq \dfrac{-c(\tilde{\matt{P}})}{a(\tilde{\matt{P}}) + b(\tilde{\matt{P}}) + d },
\end{align}
with $ a(\tilde{\matt{P}}), b(\tilde{\matt{P}}), c(\tilde{\matt{P}}) $ and $ d $ defined in \eqref{eq:appendix_a}, \eqref{eq:appendix_b}, \eqref{eq:appendix_c} and \eqref{eq:appendix_e}, respectively.

Next, we calculate the derivative of the \gls{MSE} \gls{wrt} the precoding matrix as 
 \begin{align} \label{eq:appendix_mse_derivative}
\frac{\partial \varepsilon}{\partial \tilde{\matt{P}}}  = \beta^{2}\left (\frac{\partial a(\tilde{\matt{P}})}{\partial \tilde{\matt{P}}}  + \frac{\partial b(\tilde{\matt{P}})}{ \partial \tilde{\matt{P}}}\right )  - 2\beta \frac{\partial c(\tilde{\matt{P}})}{\partial \tilde{\matt{P}}},
\end{align} with $ a(\tilde{\matt{P}}), b(\tilde{\matt{P}})$ and $ c(\tilde{\matt{P}}) $ defined in \eqref{eq:appendix_a}, \eqref{eq:appendix_b} and \eqref{eq:appendix_c}, respectively.  Closed-form expressions of the derivative terms in \eqref{eq:appendix_mse_derivative} can be written as
{\begin{align}
	\frac{\partial a(\tilde{\matt{P}})}{\partial \tilde{\matt{P}}} =&
	2\dfrac{2}{\pi}\left[
	\bar{\matt{H}}\bar{\matt{H}}^{\T} + \left (\dfrac{\pi}{2}-\dfrac{7}{6}\right )\diag{\bar{\matt{H}}\bar{\matt{H}}^{\T}}
	\right]
	\tilde{\matt{P}}\matt{R}_{\bar{\vect{s}}} 
	 \\
\frac{\partial b(\tilde{\matt{P}})}{ \partial \tilde{\matt{P}}} =	& 	2\dfrac{2}{\pi}\left[
 \dfrac{1}{2} 
	(\tilde{\matt{P}}\matt{R}_{\bar{\vect{s}}}\tilde{\matt{P}}^{\T})^{\circ 2}\circ
	\nondiag{
		\tilde{\matt{D}}_{\mathrm{opt}}^{-2}
		\bar{\matt{H}}\bar{\matt{H}}^{\T}
		\tilde{\matt{D}}_{\mathrm{opt}}^{-2}
		}
	\right .\nonumber \\
	&\left .-
	\dfrac{1}{3}
	\diag{\left(
		\tilde{\matt{P}}\matt{R}_{\bar{\vect{s}}}\tilde{\matt{P}}^{\T}
		\right)^{\circ 3}
		\tilde{\matt{D}}_{\mathrm{opt}}^{-2}
	\bar{\matt{H}}\bar{\matt{H}}^{\T}}
	\tilde{\matt{D}}_{\mathrm{opt}}^{-4}
	\right. \nonumber \\
	&\left. + 
	\dfrac{1}{2}
	\diag{
		\bar{\matt{H}}\bar{\matt{H}}^{\T}
		}
	\right]\tilde{\matt{P}}\matt{R}_{\bar{\vect{s}}}  \label{eq:db_dP}\\
\frac{\partial c(\tilde{\matt{P}})}{\partial \tilde{\matt{P}}} = 	& - 2 \sqrt{\dfrac{2}{\pi}}
	\bar{\matt{H}}\matt{\Pi} \matt{R}_{\bar{\vect{s}}}.
	\end{align}}

Finally, the projection function $ \mathcal{P}_{C}(\cdot) $ is simply defined as the normalization: $ \pcal_{C}\left(\tilde{\matt{P}}\right)\coloneqq \sqrt{E_{\mathrm{Tx}}/\tr{
		\tilde{\matt{P}}
		\matt{R}_{\bar{\vect{s}}}
		\tilde{\matt{P}}
	}
}\cdot\tilde{\matt{P}} $.


\bibliographystyle{IEEEtran}
\bibliography{IEEEabrv,Journal_1Bit_MU_MISO_2018}

\end{document}